\documentclass[nonacm,sigconf]{acmart}

\renewcommand\footnotetextcopyrightpermission[1]{} 
\AtBeginDocument{%
  }

\acmConference[ACSAC '23]{Make sure to enter the correct
  conference title from your rights confirmation email}{December 04--08,
  2023}{Austin, TX}
\acmPrice{15.00}
\acmISBN{978-1-4503-XXXX-X/18/06}

\usepackage{multirow}
\usepackage{xcolor}
\usepackage{pifont}
\usepackage{amsthm}
\newtheorem{definition}{Definition}

\usepackage{enumerate}

\usepackage{tikz}
\newcommand\encircle[1]{%
  \tikz[baseline=(X.base)] 
    \node (X) [draw=red, font=\small, minimum size=1pt, shape=circle, anchor=north, inner sep=0, scale=0.7, text=red] {\strut #1};}

\usepackage{caption}
\usepackage{subcaption}

\usepackage{float}

\usepackage{tcolorbox}
\newtcolorbox{mybox}{colback=black!4!white,
colframe=black!20!white, coltitle=black, left=0pt,
right=0pt,
top=0pt,
bottom=0pt}

\makeatletter
  \if@ACM@anonymous
    
  \fi
\makeatother

\begin{document}

\title{Assessing the Effectiveness of Binary-Level CFI Techniques}


\author{Ruturaj Vaidya}
\email{ruturaj@ku.edu}
\affiliation{%
\institution{University of Kansas}
\city{Lawrence}
\state{Kansas}
\country{USA}
}
\author{Prasad Kulkarni}
\email{prasadk@ku.edu}
\affiliation{%
\institution{University of Kansas}
\city{Lawrence}
\state{Kansas}
\country{USA}
}



\begin{abstract}
Memory corruption is an important class of vulnerability that can be leveraged to craft control flow hijacking attacks. 
\textit{Control Flow Integrity (CFI)} provides protection against such attacks. Application of type-based CFI policies requires information regarding the number and type of function arguments.
Binary-level type recovery is inherently speculative, which motivates the need for an evaluation framework to assess the effectiveness of binary-level CFI techniques compared with their source-level counterparts, where such type information is fully and accurately accessible. 
In this work, we develop a novel, generalized and extensible framework to 
assess how the program analysis information we get from state-of-the-art binary analysis tools affects the efficacy of type-based CFI techniques. 
We introduce new and insightful metrics to quantitatively compare source independent CFI policies with their \emph{ground truth} source aware counterparts.
We leverage our framework to evaluate binary-level CFI policies implemented using program analysis information extracted from the IDA Pro binary analyzer and compared with the ground truth information obtained from the LLVM compiler, and present our observations.
\end{abstract}

\maketitle

\section{Introduction}
\label{sec:introduction}

%
%
%
%
Software written in memory unsafe languages like {\tt C} and {\tt C++} is vulnerable to memory attacks. The combination of OS 
and compiler techniques such as Stack Canaries\,---\,which catch some buffer overflow attacks by checking the sanity of a random value placed on the stack prior to return, ASLR (Address Layout Randomization)\,---\,which pseudo-randomizes the memory layout of the program and, DEP (Data Execution Prevention)\,---\,which prevents execution of certain memory regions, are currently deployed in prevent these attacks.
However, {\it Code-Reuse Attacks (CRA)} such as return-into-libc (full function reuse attack)~\cite{designer1997}, ROP (Return Oriented Programming)~\cite{rop2007, checkoway2010, Snow2013} and COOP (Counterfeit Object-oriented Programming)~\cite{schuster2015, team2015bctf, Crane2015, Lan2015, Lettner2016} employ existing code segments to circumvent such defense mechanisms.
These attacks abuse vulnerabilities like buffer overflows, take control of the program and leverage gadgets to craft the control flow hijacking exploits, resulting in changes in the control flow of the program to user unintended locations.

{\it Control flow integrity (CFI)}~\cite{abadi2005} is a popular technique to prevent such control flow hijacking attacks.
CFI aims to ensure that the control flow of the program stays within the legitimate targets desired by the programmer. 
Usually, this is achieved by computing the user intended control flow targets using a static analysis phase to insert security checks into the generated binary code.
The inserted security checks monitor and enforce the control flow of the program to stay within the desired target locations at run-time. 
CFI implementations that balance security and performance have been integrated in popular compilers (e.g. {\it LLVM}~\cite{llvmcfi2022} and {\it GCC}~\cite{gcccfi2023}), making CFI accessible to ordinary programmers.
Even when CFI techniques fail to accurately determine programmer intended program control flow, they make the attackers' life difficult by constraining the control flow of the program during execution.

Various CFI techniques have been proposed after the introduction of an exemplary CFI model by Abadi et al.~\cite{abadi2005}.
CFI techniques could be source-code aware\,---\,implemented at source or compiler-level, or source-code independent\,---\,implemented at the binary level. 
Binary-level CFI techniques are necessary to secure \emph{unprotected} and \emph{untrusted} programs and third-party libraries that are typically shipped without their corresponding high-level source codes.

CFI techniques typically require the accurate recovery of function call-site and function signature information, including argument counts and all argument types.
Some advanced CFI policies also employ class and vtable hierarchy information.
For many languages like C/C++, much of this information is embedded in the source code, and can be precisely recovered by source-level tools.
Unfortunately, it becomes much harder to recover precise program structure, semantic, and other analysis information after compilation. 

Lack of accurate program analysis information at the binary-level makes it extremely challenging to build a precise function call-graph for large binary software.
In turn, the effectiveness of binary-level CFI techniques depend and suffer from the inaccuracies of the program information extracted by the adopted binary analysis framework.
Over- or under-approximation by CFI techniques can result in false negatives (attacks go undetected) or false positives (correct control flow tagged), which can dent the usability of CFI.

Our goal in this work is to investigate and quantify the correctness 
of binary-level CFI techniques and how they are impacted by the inaccuracies in program analysis information recovered by modern binary-level tools.
We focus on \emph{type-based} CFI techniques that use the number and type of arguments to match each call-site to the set of \emph{potential} call-targets.
Source-level CFI techniques have access to precise program information, and are therefore most likely to achieve their stated objective.
Therefore, we argue that the output of each source-level CFI technique can be used as the \emph{ground truth} to assess the accuracy of the corresponding  binary-level CFI technique (that operate without access to similarly accurate program information).

In this work, we develop a generalized, modular, and extensible framework, called \emph{Binary-CFI}, to study and quantify the effectiveness of different binary-level CFI techniques. 
Our framework supports the integration of different source (compilers) based and binary-level analysis modules to gather program information required to model different CFI techniques, each at the source and binary levels.
To validate our framework, we develop a source-level analysis module using the LLVM compiler~\cite{llvm2023} and a binary-level analysis module using the IDA Pro and Hex-Rays software reverse engineering (SRE) tools~\cite{idapro2022}. The analysis modules statically recover program information, including call-site and call-target argument counts, argument types, and the function return type.
We also model four different CFI techniques that employ the analysis information gathered by the source/binary-level analysis modules to impose the call-target constraints i.e. to restrict the reachable targets at each call-site to legitimate function entry points.


Next, we introduce new and insightful metrics to quantitatively compare the effectiveness of CFI policies instituted at the binary level with the \emph{ground truth} provided by their source-aware counterparts. 
Unlike most existing CFI metrics that only measure the \emph{number} of call-targets reached 
without regards to their \emph{correctness} compared to the ground truth set of call-targets~\cite{zhangsekar2013, tice2014, aia2016, Burow2017, Muntean2019, Frassetto2022}, our approach provides a more correct metric for evaluating the accuracy of binary-level CFI techniques.
We leverage our framework and metric to study and assess the effectiveness of four different CFI techniques, focusing on forward edge CFI.
We make the following contributions in this paper:
\begin{itemize}
\item [--] We develop a modular and extensible framework\footnote{Our framework is available online - https://github.com/Ruturaj4/B-CFI} along with a common language to compare the accuracy and effectiveness of 
binary-level type-based CFI techniques.
\item [--] We develop a mechanism to model multiple different type-based CFI techniques using program information obtained from different sources.
\item [--] We develop metrics to quantitatively measure the accuracy of binary-level CFI techniques compared to \emph{ground truth} results obtained with access to the source code.
\item [--] We compare the accuracy of CFI relevant analysis 
recovered by modern SRE tools (with and without symbol information) compared to that extracted by standard compilers.
\item [--] We employ our framework, models, metrics and mechanisms to recover program information from IDA Pro, a state-of-the-art binary analyzer, and LLVM, a modern compiler, and employ that information to quantitatively assess the accuracy of four binary-level CFI techniques compared to their source-level equivalents. We present our evaluation results and discuss our observations and takeaways.
\end{itemize}


\section{Background and Related Work}
\label{background}
In this section we describe technical background and related research in CFI techniques, tools and metrics, and binary analysis.

\subsection{Control Flow Integrity}
Code-Reuse Attacks (CRA)~\cite{designer1997, rop2007, checkoway2010, Snow2013, schuster2015, team2015bctf, Crane2015, Lan2015, Lettner2016} allow attackers to exploit spacial and temporal memory safety violations to alter the control flow of the program. {\it Control Flow Integrity (CFI)} provides promising protection against such arbitrary control flow subversion. CFI techniques use static or dynamic analysis to compute the program control flow graph (CFG) and then check if the program execution follows the CFG computed in the previous analysis stage at run-time. Thus, CFI maintains program integrity by allowing only legitimate control transfers during execution.

\begin{figure}[h]
    \centering
    \includegraphics[width=1.0\columnwidth]{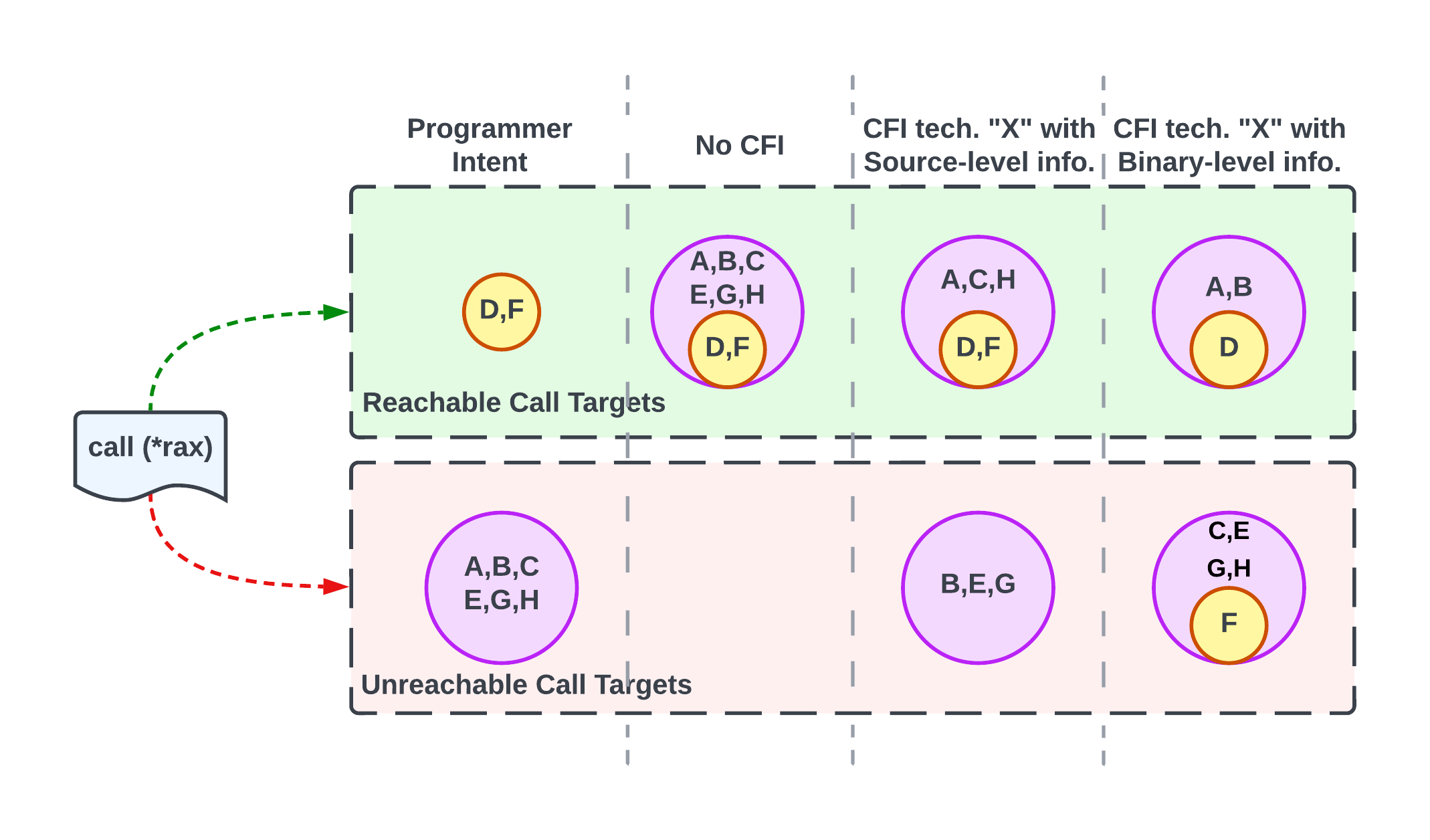}
    \caption{High-level overview of CFI techniques}
    \label{fig:source_vs_bin}
\end{figure}

We use Figure~\ref{fig:source_vs_bin} to describe, at a high-level, how CFI techniques work and also to explain our goals in this work.
When coding, the developers most likely intend the control-flow at each \emph{indirect call-site} to only reach a few potential function \emph{targets} during program execution.
Thus, the programmer \emph{intent} in Figure~\ref{fig:source_vs_bin} is for the call-site to only reach targets `{\tt D}' and `{\tt F}'. 
Unfortunately, this programmer intent is not explicitly encoded in the source-code, and is lost before it reaches the compiler.
Without any CFI check, an attacker may be able to subvert the call-site to reach \emph{any} reachable function target (`{\tt A}', `{\tt B}', `{\tt C}', `{\tt D}', `{\tt E}', `{\tt F}', `{\tt G}', and `{\tt H}' in Figure~\ref{fig:source_vs_bin}).

Different CFI techniques use various \emph{safe} approaches that constrict the set of spurious reachable targets, while ensuring that the technique does not inadvertently disallow any correct (but, unknown) programmer-intended targets.
If a correct target is not in the set of reachable targets, then the CFI check may trigger a \emph{false positive} alarm for correct program flow during execution.
At the same time, if the set of reachable targets is too broad, then the CFI technique may leave the program more vulnerable to attacks.
The source-level CFI technique in Figure~\ref{fig:source_vs_bin} has been \emph{designed} to partition the targets into reachable and unreachable sets, as illustrated.

Unfortunately, program analysis information recovered by binary-level SRE tools may be imprecise, which can cause the \emph{same} CFI algorithm to produce different and incorrect reachable and unreachable target function sets at the binary-level for each call-site (as illustrated in Figure~\ref{fig:source_vs_bin}).
Our goal in this work is simply to measure and study this imprecision in the output of binary-level CFI techniques as compared to their source-level counterparts.
\footnote{Consequently, even if the binary-level CFI technique produces a more desirable outcome (for example, by allowing all programmer-intended targets and a smaller spurious set in the reachable set), it will still be considered erroneous in this work, if it does not exactly match the output of the corresponding source-level approach, since the technique did not function as algorithmically designed (due to imprecise input analysis information), and any observed ``\emph{improvement}'' is merely coincidental.}



Abadi et al. introduced the idea of CFI by statically computing the CFG and restricting control flow of the program to the valid targets during run-time~\cite{abadi2005}. 
Since then, researchers have developed many CFI policies and algorithms that differ in their implementation, precision and cost.
Several CFI approaches employ pointer analysis to construct the CFG that is needed by the algorithm~\cite{zhangsekar2013, zhang2013, tice2014, wang2015}.
However, static points-to-analysis is imprecise, especially for program binaries~\cite{Farkhani2018}. 
Therefore, researchers have proposed CFI techniques that incorporate program invariants such as argument count and types to construct the CFG.
These type of techniques are referred to as Run-time Type Checking (RTC) based CFI techniques~\cite{tice2014, mcfi2014, rap2015,veen2016, Muntean2018, Farkhani2018, llvmcfi2022}.

In this work, we do not propose or build new CFI techniques.
Instead, we develop a new framework and metrics to model and compare binary-level RTC based CFI mechanisms against a known ground-truth.
We also assess the accuracy of the relevant program information recovered by state-of-the-art binary analysis tools, and their impact on the precision of binary-level CFI policies.

\subsection{CFI Security Policy Comparison Metrics}
Researchers have developed several mechanisms and metrics to evaluate and compare the protection provided by different CFI policies.
{\it Average Indirect target Reduction (AIR)}~\cite{zhangsekar2013} measures the reduction of permitted call-targets.
{\it AIA}~\cite{aia2016} computes the average number of call-targets per function call. 
Similarly, {\it fAIR}~\cite{tice2014} and {\it fAIA}~\cite{Frassetto2022} are forward-edge variations of the previous metrics.
The {\it CTR (Call-Target Reduction)} metric provides absolute values (rather than averaged results) of reachable call-targets at every indirect call-site~\cite{Muntean2019}. 
Most of these metrics use a relative measure, such as reduction in the average number of reachable \emph{targets} from each call-site, or reduction in the number potential gadgets, etc. to assess the accuracy and benefit of the CFI technique.

Burow et al. 
propose a metric called {\it QuantitativeSecurity} that computes the number of equivalence classes and the inverse of the size of the largest class, to quantify the security of CFI techniques~\cite{Burow2017}. 
Frassetto et al. develop the {\it BLOCKInsulation} and {\it CFGInsulation} metrics to calculate the distance between a vulnerable instruction to system call at basic-block granularity~\cite{Frassetto2022}. 

None of these existing CFI metrics incorporate the notion of obtaining the actual accuracy of any CFI technique as compared to some known ground truth, and determining the false positive and false negative call-targets at each call-site. 
In this work, we show why such earlier CFI metrics are ill-suited for comparing the performance of binary-level CFI policies.
We introduce new metrics that can quantitatively compare the accurate call-targets in each equivalence class identified by binary-level policy with that of call-targets recuperated in the corresponding equivalence class using a source aware ground truth policy.



\subsection{CFI Frameworks}
It is difficult to compare and assess the security of different CFI policies as researchers use different settings, including compilers, flags, and operating systems, to implement their techniques. 
Therefore, researchers have built detailed frameworks, mechanisms, and metrics to compare and assess CFI techniques uniformly.

Farkhani et al. develop a framework to analyze the ability of RTC CFI mechanisms, and compare them with a points-to analysis based CFI mechanism~\cite{Farkhani2018}.
Li et al. introduce CScan\,---\,a framework to compute actual feasible targets using run-time checks and CBench\,---\,an extensive set of vulnerable programs to assess the effectiveness of CFI techniques~\cite{cracks2020}.
{\it ConFIRM}~\cite{confirm2019} analytically compares various CFI policies in terms of compatibility issues in contrast to focusing on performance or security. 

Our framework to evaluate binary-level CFI policies is inspired by a compiler-level CFI policy comparison framework, called LLVM-CFI~\cite{Muntean2019}.
This framework provides a LLVM-Clang based unified framework for statically modelling and systematically assessing various CFI techniques. 
LLVM-CFI leverages a link time optimization ({\tt LTO}) pass in the LLVM compiler to impose constraints on invariants collected during compilation to implement CFI policies. 
The CFI policies in our current work also adhere to much of the formalization described by this earlier work. 
However, our goals, implementation machinery and metrics used differ considerably from LLVM-CFI.

The CFI policies in LLVM-CFI are modelled based on their idealized representation, which means that they do not consider the effect of loss in high-level information whilst modelling source insentient CFI techniques.
In other words, the binary-level CFI policies in LLVM-CFI are established on the premise that the analysis primitives are all recovered correctly at the binary level. 
Instead, our goal in this work, which is to compare the precision of binary-level CFI policies, requires us to gather the necessary program information from both binary-level and source-level analysis tools.


None of these earlier CFI policy comparison frameworks and metrics attempt to study and assess how the loss of program information at the binary-level affects the efficacy of different binary-level CFI policies compared to some ground truth, which we do in this work.
Additionally, we also develop a new set of metrics that can more accurately determine the accuracy of binary-level CFI policies compared to a ground truth, which was not attempted by earlier CFI policy comparison frameworks.


\subsection{Binary Analysis}
Many previous research works have attempted to study and resolve the precision of binary-level static analysis tools.
Researchers have compared and systematically studied various disassemblers in terms of accuracy, and the algorithms or heuristics employed in different SRE tools~\cite{Andriesse2016, pang2020}. 
Other works identify code structures that make it challenging for reverse engineering tools to properly disassemble binary code and construct a correct control flow graph (CFG)~\cite{meng2016}. 
Researchers have also studied the ability of binary analysis frameworks to recover types or challenges of type recovery at binary level~\cite{noonan2016, caballero2016}. 
Liu and Wang 
assess commercial and open source decompilers in terms of usability and effectiveness~\cite{liu2020}. 

These previous techniques attempt to understand the challenges in binary analysis.
However, they do not assess the impact of the imprecision in the collected analysis information on the accuracy of type-based binary-level CFI techniques, like we do in this work.

\section{Implementation}
\label{implementation}
In this section we describe the design and implementation of the CFI policy comparison framework and the CFI policy models that we built and used for this work.

\subsection{Design Overview}

\begin{figure}[h]
    \centering
    \includegraphics[width=1.0\columnwidth]{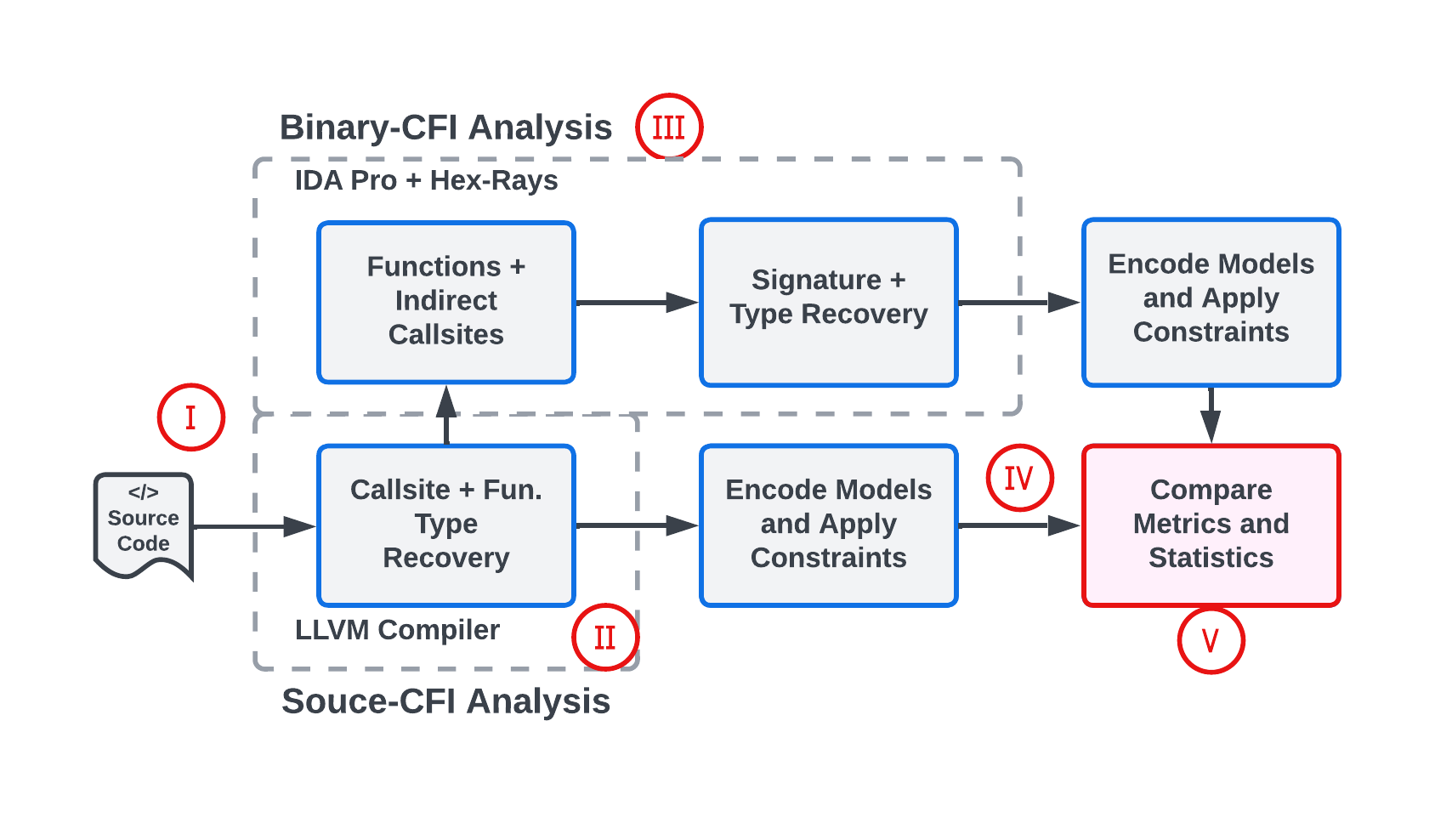}
    \caption{Block Diagram of Binary-CFI}
    \label{fig:block}
\end{figure}

In this paper, we introduce {\it Binary-CFI}\,---\,a binary level CFI comparison framework. 
To assess and analyze the precision of type-based binary-level CFI techniques, we design and construct an evaluator framework that is capable of comparing the results achieved by different CFI techniques at both the source-sentient and insentient levels. 
Figure~\ref{fig:block} shows the high-level block diagram of our evaluation technique. 
The technique can be broadly classified into two stages. Firstly, relevant program analysis information is collected from both source-level (LLVM) and binary-level (IDA Pro) means and secondly, this information is fed into the CFI models, and the results are computed, compared and analyzed.

In further detail, our technique performs the following steps:
\encircle{I} First, the source code to be analyzed is compiled (using the LLVM compiler, in this work). 
\encircle{II} During compilation, we collect various program analysis information, including function argument counts, and argument and function types at each call-site and at every call-target using a dynamically loadable {\it LLVM LTO} (Link Time Optimization) pass that we built for this work. This pass makes separate compilation of source files possible providing flexibility.
These source-level analysis statistics are
used to drive an idealized representation (or ground truth) of our type-based CFI policies.

\encircle{III} The output binary is then employed for Binary-CFI statistics collection. 
In this work we leverage IDA Pro~\cite{idapro2022}\,---\,a popular reverse engineering framework to statically analyze the binaries and recuperate static analysis information, including indirect calls and program functions accompanied by their type signatures i.e. function return type, function argument counts and their types at each call-target and call-site. 
We also leverage Hex-Rays decompiler to refine the type information generated by IDA Pro. 
The advanced type inference in the decompiler assists us to model robust run-time type checking (RTC) policies at the binary level.

We invoke our LLVM LTO pass after full link time optimizations to ensure the accurate source to binary function matching. We do not consider any unmatched functions if they aren't identified correctly by IDA tool.
Although we employ LLVM and IDA Pro for this work, our framework is modular and evaluators can use any other source- and binary-level static analysis tools  to extract function and call-site related program analysis information.

\begin{figure}[t]
    \centering
    \includegraphics[width=0.5\columnwidth]{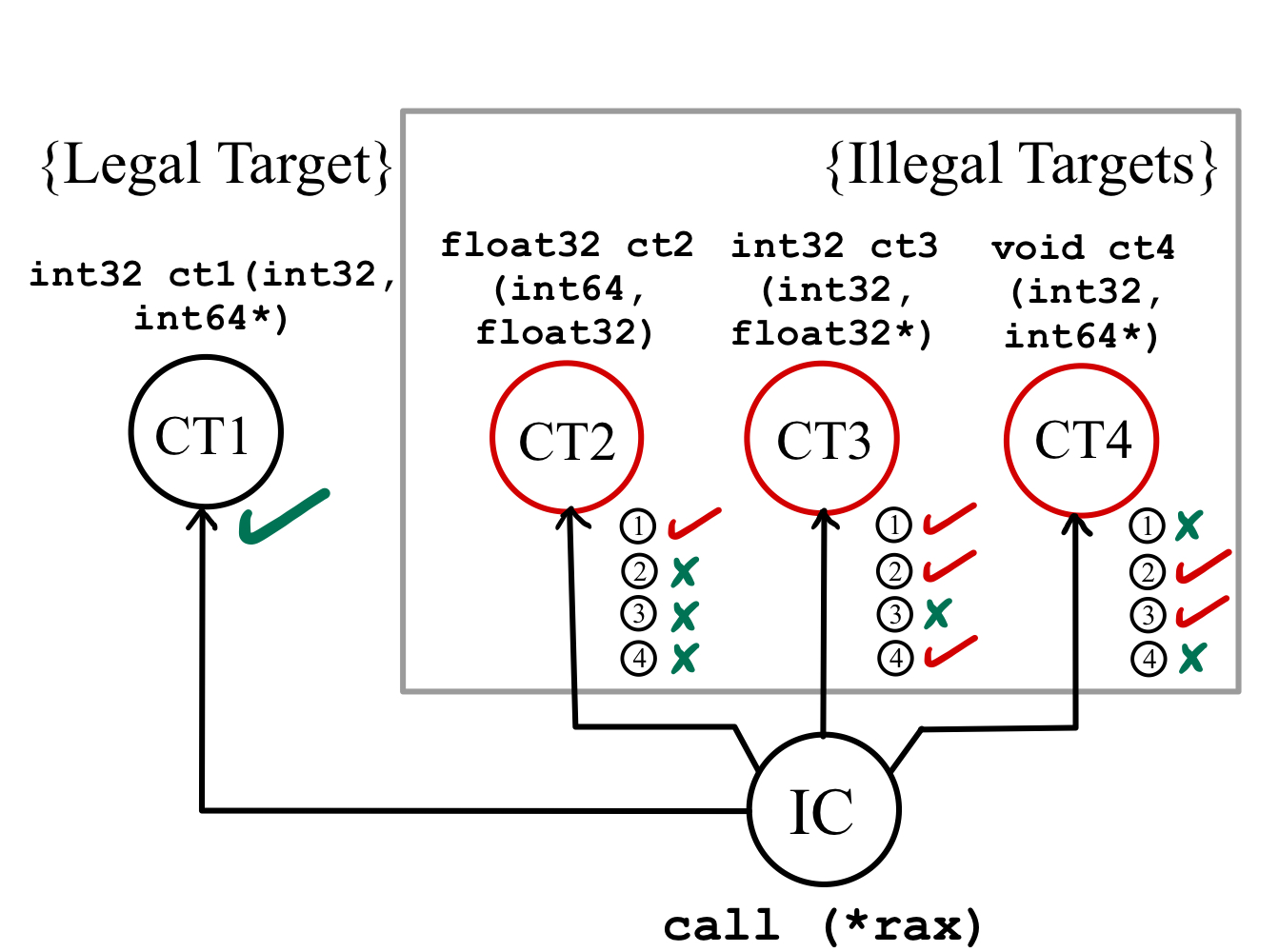}
    \caption{Indirect call-site targeting functions in binary hardened with four different policies\,---\,\ding{172} TypeArmor, \ding{173} IFCC, \ding{174} MCFI and \ding{175} $\tau$CFI}
    \label{fig:call-site}
\end{figure}

\encircle{IV} After the recovery of these analysis primitives at both the source level and binary-level, type-based policy constraints are applied corresponding to each deployed CFI policy. 
At this stage, evaluators can select and encode any CFI policy of their choice by setting various type-based constraints. 
Thus, this extensible and convenient framework will enable analysts to implement and verify new type-based CFI policies at the binary level without doing repetitive compilation and analysis. 
To validate our framework, we implement and deploy four type-based policies (explained in details in \ref{type_based_constraints}) for evaluation. 

\encircle{V} Finally, the output of the CFI models using source-level and binary-level program information is compared and the final results are displayed to the evaluator.

\subsection{Type-Based CFI Policies}
\label{type_based_constraints}

In this section we describe the four type-based CFI policies we model by applying different type-based constraints. 
Some of these were also used and compared in the LLVM-CFI work~\cite{Muntean2019}. 

Figure~\ref{fig:call-site} displays an indirect call-site targeting four different functions in a binary hardened by modeling four different type-based CFI policies. 
The function shown on the far-left ({\tt CT1}) is the only legal call-target intended to be called from indirect call (IC) instruction {\tt call (*rax)}. Besides, three other functions ({\tt CT2}-{\tt CT4}) are illegal call-targets and should ideally be unreachable during correct program execution. We assume that the attacker controls the value of register {\tt rax}.

We now discuss constraints and type collisions imposed by the four CFI policies we employ. However, our technique is adaptable and evaluators can introduce and model other policies with various levels of type-based precision.
\\\\
{\bf \ding{172} TypeArmor}~\cite{veen2016} was originally implemented at the binary level by using coarse-grained type invariants. The policy considers the number of arguments without explicit types. At each call-site the call is allowed only if the number of arguments at the call-target are equal or less than that at the call-site (maximum up to six). Additionally, void and non-void functions are differentiated i.e. call-sites which expect a return value must only target functions with non-void return type. Note that such assumptions can not be made on the contrary, i.e. if a call-site doesn't expect a return value, then it can call void as well as non-void functions. This relaxed policy is practical at the binary level, as it is often difficult to infer whether the function is going to return a value or not. Thus, at the example call-site in Figure~\ref{fig:call-site}, the TypeArmor CFI policy allows the {\tt call (*rax)} instruction to reach {\tt CT1}, {\tt CT2} and {\tt CT3} functions, which includes two illegal targets.
\\\\
{\bf \ding{173} IFCC}~\cite{tice2014} is implemented similar to the encoding explained in \cite{Muntean2019}. IFCC takes into account the argument and parameter counts, along with their basic types to match call-sites to call-targets. However, base pointers types are not considered, i.e. {\tt void*} and {\tt int*} are considered equivalent. Therefore, functions {\tt CT3} and {\tt CT4} in Figure~\ref{fig:call-site} are allowed (in addition to {\tt CT1}). Return type is not taken into consideration. Note that the types are not over-approximated i.e. they are not considered as upper bound, but are matched according to the exact type.
\\\\
{\bf \ding{174} MCFI}~\cite{mcfi2014} is a CFI policy that is stricter than IFCC in terms of how pointer types are recuperated. Pointer types such as {\tt void*} and {\tt int*} are considered distinct. Similar to IFCC, the number of parameters and their types are matched with call-site argument count and types. However, stricter types are taken into consideration. Thus, as seen in Figure~\ref{fig:call-site} only one target i.e. {\tt CT4} is reachable with the stricter MCFI policy (in addition to {\tt CT1}). Function return types are not considered, similar to IFCC.
\\\\
{\bf \ding{175} $\tau$CFI}~\cite{Muntean2018} considers argument and parameter types along with their counts. The types are contemplated based on the size of the registers {\tt \{0,8,16,32,64\}} prepared during the indirect call. According to x86-64 calling convention (System V ABI) the first 6 arguments are passed through registers during a function call. $\tau$CFI policy allows the call if 1) the number of arguments prepared at the call-site are more or equal to the number of parameters consumed at the call-target, 2) the return type recuperated at the call-site and the call-target is non-void and its size at the call-site is larger than that of the call-target return type; else, if return type recuperated at the call-site is void and then it can also call non-void functions, 3) the size of the argument types at call-site are greater than or equal to their matching arguments at call-targets. Thus, in our example displayed in Figure~\ref{fig:call-site}, {\tt CT3} is the only illegal target that is allowed to be reached from the indirect call-site.

\section{Evaluation}
\label{evaluation}
In this section we present our experimentation framework and benchmarks. We also introduce and describe our novel evaluation metrics, and discuss our experimental results.

\subsection{Benchmarks}
We evaluate our framework using {\it sixteen} {\tt C} and {\tt C++} benchmarks from the SPEC 2006 integer and floating point suite. We leave out the remaining benchmarks either because we didn't find any indirect call-sites in the optimized benchmark version (mcf, libquantum and lbm) or when the benchmarks use {\tt Fortran} code.

Additionally, we include five popular and large real world applications for this study.
Specifically, we performed our evaluation with 
(a) {\it Nginx} (v1.22.1 {\tt C}), an open-source web server software, 
(b) {\it Node JS} (v10.24.0 {\tt C/C++}), an open-source, cross-platform JavaScript run-time environment, 
(c) {\it Apache Traffic server} (v6.2.3 {\tt C/C++}), an open-source forward and reverse proxy web server, 
(d) {\it postgresql} (v12.0 {\tt C}), an open-source relational database management framework, and the
(e) Tor Browser (v0.4.8.0-alpha-dev {\tt C}), an open-source web browser focused on privacy and security. 
We obtained the most primary application binary from these benchmarks for our analysis.

Our benchmarks along with the total number of indirect call-sites and call-targets in each program are listed in Table~\ref{tab:benchmark_properties}. All the SPECint and SPECfloat benchmarks are presented together in their respective groups in this table (and in all later results).

\subsection{Experimental Configuration}
We design two benchmark configurations for this study.
\begin{enumerate}[I.]
    \item {\bf Ideal or Baseline Scenario:} For our first configuration, we keep the debugging symbols and compile the binary with optimizations (`{\tt -O3}'). We refer to this configuration as the \emph{baseline}. This \emph{baseline} configuration can be considered as an idealized representation at the binary level where some source semantics in the form of debug symbols are available to guide the binary analysis frameworks. 
    \item {\bf Practical Scenario.} For our second configuration, we strip the debugging symbols using `{\tt GNU strip}'. This is a practical scenario for most COTS (Commercial off-the-shelf) binaries and presents a more challenging case for the binary analysis algorithms. All benchmarks are still optimized by `{\tt -O3}'.
\end{enumerate}

Our framework computes the number of allowed target functions for each indirect call-site according to the constraints imposed by each respective CFI policy. 
We extract function and call-site signatures, including the number of parameters/arguments and their types using an out-of-tree LLVM pass to collect the ground truth program analysis information.
Likewise, we extract equivalent information about the program binary using the IDA Pro reverse engineering framework.

All experiments are performed on Fedora 34 operating system with x86-64 Intel Xeon processor. The LLVM/Clang version used is (v.12.0.0) to compile binaries and get the ground truth program information, and 64-bit version of {\it IDA Pro} (v7.5.2) is used to conduct binary analysis and extract the program information used by the binary-level CFI models. 


\begin{table}[]
    \caption{Benchmark Properties}\label{tab:benchmark_properties}
    \centering
    \resizebox{0.5\columnwidth}{!}{
    \begin{tabular}{c|c|c}
        \hline
        Benchmark & call-targets & call-sites \\
        \hline
        SPECint & 15594 & 20304\\
        SPECfp & 2341 & 1179\\
        nginx & 1237 & 448\\
        postgresql & 11089 & 9367\\
        trafficserver & 6886 & 8311\\
        tor & 5761 & 273\\
        node & 133496 & 8239\\
        \hline
    \end{tabular}
    }
\end{table}


\subsection{Evaluation Metric}
\label{rctr_metric}

To compare and evaluate the \emph{precision} of binary-level CFI policies with their source aware counterparts in terms of the \emph{correct} reachable call-targets at each call-site, we introduce new metrics that calculate not only the number of targets reached (fewer the better), but also employ the known ground-truth targets information to check if there are any false positives or false negatives generated by the CFI policy under evaluation.
Such a detailed evaluation of CFI policies is crucial, as mere call-target reduction results, as measured by most earlier CFI metrics, can not characterize the number of:
\begin{itemize}
\item true positives -- illegal (unreachable) targets that are correctly marked by the CFI policy under evaluation,
\item false positives -- legal (reachable) targets in the ground truth, but are marked as illegal by the CFI policy under evaluation,
\item true negatives -- legal targets in the ground truth that are correctly marked by the CFI policy under evaluation, and
\item false negatives -- targets illegal in the ground truth that are incorrectly marked as legal by the CFI policy. 
\end{itemize}
Thus, it is very important to know the exact targets reached, i.e. we not only need to check how many functions are reached using {\it Binary-CFI}, but also how many of these functions match the functions detected using our ground truth.

We introduce new metrics named $RelativeCTR_T$ and $RelativeCTR_F$ to check whether the actual targets reached when Binary-CFI policies are applied are in fact equivalent to the actual targets reached when Source-level CFI policies are applied. $RelativeCTR_T$ (higher the better) represents the number of call-targets that are accurately reached at a particular call-site using Binary-CFI policy, compared to source-level CFI policy, and $RelativeCTR_F$ (lower the better) presents the call-targets that are incorrectly reached at a particular call-site using binary-level policy, compared to its source aware CFI policy counterpart.

Suppose that $P$ is a program with total indirect call-sites $IC$ and total
reachable call-targets $CT$. Let $IC_i$ be an indirect call-site in program $P$ with number of reachable call-targets $CT_i$ after applying the CFI constraints for source aware policy $P_c$ and $CT_{i}^{'}$ be number of reachable call-targets after applying source independent policy $P_{c}^{'}$ at the same call-site. Then, $RelativeCTR_T$ and $RelativeCTR_F$ are defined as follows.

\begin{definition}
$RelativeCTR_T$ is the ratio of the intersection of targets in Source-CFI ($CT_i$) and in Binary-CFI ($CT_{i}^{'}$) to the total number of actual targets in Source-CFI ($CT_i$) at an indirect call-site $ICi$.
\end{definition}

\[RelativeCTR_T ~~(R_T)  = \sum_{i=1}^{n} {(CT_i\cap CT_i^{'})}/{CT_i}\]

\begin{definition}
$RelativeCTR_F$ is the ratio of the total number of call-targets in ($CT_{i}^{'}$) reachable with Binary-CFI but not reachable with Source-CFI ($CT_i$) to the total number of targets in Binary-CFI ($CT_{i}^{'}$) at an indirect call-site $ICi$.
\end{definition}

\[RelativeCTR_F ~~(R_F) = \sum_{i=1}^{n} {(CT_i^{'}\setminus CT_i})/{CT_i^{'}}\]

We illustrate our new metrics using the simple hypothetical example from Figure~\ref{fig:source_vs_bin}.
This program has eight different functions `A', `B', `C', `D', `E', `F', `G', and `H'.
For some indirect call-site $IC_1$ in the program, the set of reachable targets as identified by the source-level CFI policy (our ground truth) are `A', `C', `D', `F' and `H' ($CT_{1}$, $CT_{3}$, $CT_{4}$, $CT_{6}$ and $CT_{8}$). 
However, the binary-level CFI policy under evaluation determines the reachable set of targets from the same call-site to be `A', `B' and `D' ($CT_{1}^{'}$, $CT_{2}^{'}$ and $CT_{4}^{'}$).
Thus, according to the desired algorithm, `B' is an unintended target, and `C', `F' and `H' are correct targets that are missed. 
Therefore, the $RelativeCTR_T$ for this call-site is $2/5$, which indicates the correctly detected, or true negative targets (and, correspondingly, also the false positive targets).
$RelativeCTR_F$ is $1/3$, which indicates the incorrectly detected or the false negative call-targets.
Thus, a high $RelativeCTR_T$ indicates a high true negative (and low false positive rate for the CFI technique), i.e., a low likelihood of throwing a fault when there is none. 
Likewise, a high $RelativeCTR_F$ indicates a more relaxed CFI policy and a higher likelihood for the CFI technique to allow unsupported control flow paths that can lead to attacks.

In addition, since we target the same weakness in all previous CFI metrics, we use only one, the popular CTR metric~\cite{Muntean2019}, as representative of the category of metrics that only use the measure of reduction in the number of call-targets from each call-site to rate different CFI policies.
The CTR metric depicts the absolute values of the number of call-targets accessible from a call-site after hardening with a particular CFI policy. The CTR metric is defined as follows.
\[CTR = \sum_{i=1}^{n} ct_i\]
Where $ct_i$ is number of call-targets reachable from an indirect call-site $ic_i$. 
A lower value of CTR implies a better CFI policy, as it ostensibly reduces the number of \emph{extraneous} targets allowed from a call-site.
In this paper, we highlight some important shortcomings of the CTR (and similar) metrics for our work. 
Specifically, such metrics do not fairly and accurately assess the precision of CFI policies compared to some known \emph{ground truth}. 

\subsection{CFI Policy Comparison}
We present and discuss our results in this section. 
We use our framework and models to collect the $RelativeCTR$ and $CTR$ numbers for all our benchmark programs.
We use {\it Dwarf} symbols at every call-site to match the call-sites detected during the source-level LLVM pass with the call-sites in the binary executable. We leverage the {\it llvm-symbolizer} tool to match dwarf symbols with the address of the respective call-site in the binary. 
Note that the binary address to dwarf mapping is one-to-many and thus we consider all the call-sites that appear in the binary for each source-level call-site. 

We leverage our new $RelativeCTR$ metrics (the $RelativeCTR_T$ and $RelativeCTR_F$ metrics introduced earlier in Section~\ref{rctr_metric}) to show correctly and incorrectly reachable call-targets at each call-site. 
Table~\ref{table:spec-opt-stripped:rctr_short} shows the $RelativeCTR$ metrics with "Mean" values for our benchmarks in both our binary configurations {\bf I} and {\bf II}~\footnote{Detailed distribution of $RelativeCTR$ and $CTR$ metrics with Min., Max., Median, 90$^{th}$ percentile, Mean, and Standard deviation values is shown in Figures~\ref{table:spec-opt-stripped:rctr} and \ref{table:spec:ctr} in Appendix~\ref{appendix:extended_results}.}.
As mentioned earlier, the $RelativeCTR_T$ results display the number of \emph{true negative} targets (according to the ground truth) that are reached by the binary-level CFI models, while the $RelativeCTR_F$ results show the ratio of call-targets that are \emph{incorrectly} reached by the binary-level CFI algorithms.

The results in Table~\ref{table:spec-opt-stripped:rctr_short} allow us to make some important observations that would be missed by earlier CFI comparison metrics
that use a reduction in the number of reachable targets from each call-site as the only measure to evaluate the effectiveness of CFI techniques~\cite{zhangsekar2013, tice2014, aia2016, Burow2017, Muntean2019, Frassetto2022}. 
As mentioned earlier, to better understand and distinguish the merits, differences and trade offs of our new proposed $RelativeCTR$ metrics and the older class of CFI comparison metrics, we use the {\it CTR (Call-Target Reduction)}~\cite{Muntean2019} as a representative metric from the older category.

Table~\ref{table:spec:ctr_short} presents the results using the CTR metric for 4 CFI policies - \ding{172} TypeArmor, \ding{173} IFCC, \ding{174} MCFI and \ding{175} $\tau$CFI, and for our benchmark set when using the analysis information from LLVM (\emph{Source-CFI}), and our two binary configurations,
\emph{Binary-CFI} ({\bf I}) and \emph{Binary-CFI} ({\bf II}), respectively.
The CTR metrics in Table~\ref{table:spec:ctr_short} present the absolute values of reachable targets. 

Now, we employ the $RelativeCTR$ and $CTR$ numbers for our benchmark programs, presented in Tables~\ref{table:spec-opt-stripped:rctr_short} and \ref{table:spec:ctr_short}, respectively to make the following observations.
Of the four CFI policies modelled in this work, TypeArmor is the most \emph{permissive}, since it only considers argument counts and discards argument type information.
By contrast, MCFI is most \emph{strict} as it considers both basic types and mature pointer types.
Accordingly, we can see higher CTR numbers across the board for TypeArmor and lower relative CTR numbers for MCFI, which confirms this property about the CFI policies.

For this work though, it is more pertinent to compare the Binary-CFI CTR numbers with the corresponding Source-CFI numbers to assess the accuracy of CFI methods at the binary-level (by considering the Source-CFI numbers as the ground truth in each CFI case).
When using the CTR metric, the difference between the binary-level and source-level numbers indicates the potential error in the binary-level CFI models.
We find that,
\begin{mybox}
\textit{\textbf{Observation 1.} 
The binary-level CTR numbers differ significantly from the source-level CTR metrics for all our CFI models.
Furthermore, this difference is greater for the more restrictive CFI policies.}
\end{mybox}
Thus, CFI policies, like MCFI and IFCC, that rely on more precise program data type information appear to be more erroneous as compared to the simpler CFI models, like TypeArmor and $\tau$CFI.
This is an intuitive result as it indicates that errors in correctly reconstructing the type information at the binary level negatively impacts the algorithms employing such data during their computations.

While this observation derived from the CTR results appears to be correct, a deeper analysis reveals critical issues and misleading outcomes. 
For instance, the results in Table~\ref{table:spec:ctr_short} also show that the Binary-CFI CTR ratios are often tighter that the Source-CFI numbers.
We find that in 3 of the 7 benchmark categories with TypeArmor, and in all of the 7 benchmark categories with IFCC, MCFI, and $\tau$CFI, the number of \emph{mean} reachable targets from each call-site is smaller with Binary-CFI ({\bf I}) compared with the Source-CFI numbers.
This result with the CTR metric is confusing since it suggests that the binary-level techniques achieve better effectiveness with fewer extraneous call-targets compared to the source-level techniques.
Likewise, in many cases, especially for the stricter CFI policies, we can observe that the CTR numbers are tighter with the \emph{stripped} benchmarks in the Binary-CFI ({\bf II}) configuration, compared with the Binary-CFI ({\bf I}) configuration, which is again a confusing and likely misleading outcome.


Results with our new $RelativeCTR$ metric in Table~\ref{table:spec-opt-stripped:rctr_short} can help resolve this confusion caused when looking solely at the CTR numbers in Tables~\ref{table:spec:ctr_short}.
Results derived by using our new metric enable us to make several interesting observations.
\begin{mybox}
\textit{\textbf{Observation 2.} 
All CFI models for the binary configurations display high error rates that is not captured by the existing metrics used to measure the performance of CFI policies, like CTR.
}
\end{mybox}
Thus, our novel $RelativeCTR$ metrics reveal that the tighter CTR numbers with the binary-level CFI models are not a result of only eliminating the extraneous or false negative call-target edges for each call-site.
Rather, the lack of precise program analysis information at the binary-level causes the CFI models to produce significant numbers of \emph{false positive} (indicated by $RelativeCTR_T$) and \emph{false negative} (indicated by $RelativeCTR_F$) edges.

\begin{mybox}
\textit{\textbf{Observation 3.} 
Binary-level CFI models, like MCFI and IFCC, that rely on more precise program analysis information are more erroneous, compared to the simpler CFI models, like TypeArmor and $\tau$CFI.
}
\end{mybox}
We find that the \emph{Mean} $RelativeCTR_T$ values are significantly lower for all benchmark categories with the stricter MCFI  and IFCC CFI policies compared to TypeArmor and $\tau$CFI. 
Likewise, the \emph{Mean} $RelativeCTR_F$ values are much higher for MCFI and IFCC compared to TypeArmor and $\tau$CFI.
While this is not a particularly surprising result in hindsight, the extent of the observed error is quite staggering.
Thus, we find that the \emph{mean} number of \emph{correct} or \emph{true negative} ($RelativeCTR_T$) edges recovered by the MCFI policy even in the Binary configuration {\bf I} (with debug symbols available) drops to as low as 0.10 and 0.14 for the {\tt trafficserver} and {\tt SPECint} benchmark categories, respectively, and with less than 50\% of the \emph{true negative} edges recovered for all but one benchmark suite.
Likewise, the number of \emph{incorrect} or \emph{false negative} ($RelativeCTR_F$) edges recovered is as high as 0.66 and and 0.65 with the MCFI policy in the Binary configuration {\bf I} for benchmark suites {\tt postgresql} and {\tt SPECint}, respectively.
It is also interesting to note that binary-level SRE tools struggle to recover precise program analysis information even for binaries with debug symbol information available, resulting in poor performance by CFI models employing such information.
This level of imprecision by binary-level CFI techniques is not something that has been observed or reported by earlier works that used simple metrics like the CTR.

\begin{mybox}
\textit{\textbf{Observation 4.}
Binary-level CFI policies produce significantly more erroneous results for benchmarks that are \emph{stripped} of debugging symbols, compared to binaries that retain their debug symbols information.
}
\end{mybox}
From Figure~\ref{table:spec-opt-stripped:rctr_short} we can also observe that in almost every case, the $RelativeCTR_T$ values are lower, while the $RelativeCTR_F$ values are higher for benchmarks that have been stripped of debug symbols (binary configuration {\bf II}) compared to programs with debug information intact (configuration {\bf I}).
Thus, it is clear that the greater imprecision in static analysis information that is recovered by SRE tools for stripped binaries results in degrading the performance for security and optimization algorithms that rely on such data.
While this is also an expected and intuitive result, there has never previously been an attempt or a mechanism to observe, measure, and report the  amount of error in CFI policies.
If anything, it is interesting to note that the magnitude of error displayed by binary-level CFI policies in configuration {\bf II} programs, with symbols stripped, is not very large in several cases, compared to the inherent error already present in CFI models in configuration {\bf I}.
We even find that in a few cases, like the \emph{mean} $RelativeCTR_T$ for {\tt trafficserver} with the MCFI policy, and the \emph{mean} $RelativeCTR_F$ for {\tt postgresql} with the IFCC policy, stripped benchmarks produce marginally better performance compared with unstripped benchmarks.

\begin{table*}[]
    \caption{Mean $RelativeCTR$ comparison results of our 4 CFI policies ({\it TypeArmor}, {\it IFCC}, {\it MCFI} and {\it $\tau$cfi})}\label{table:spec-opt-stripped:rctr_short}
    \centering
    \resizebox{1.5\columnwidth}{!}{
    \begin{tabular}{|c|||r|r|r|r||r|r|r|r||r|r|r|r||r|r|r|r|}
         \hline
         \multicolumn{1}{|c|||}{} & \multicolumn{4}{|c||}{TypeArmor}
         & \multicolumn{4}{|c||}{IFCC} & \multicolumn{4}{|c||}{MCFI}
         & \multicolumn{4}{|c|}{$\tau$CFI}\\
         \hline
        Benchmark & $R_T$ (I) & $R_F$ (I) & $R_T$ (II) & $R_F$ (II) & $R_T$ (I) & $R_F$ (I) & $R_T$ (II) & $R_F$ (II) & $R_T$ (I) & $R_F$ (I) & $R_T$ (II) & $R_F$ (II) & $R_T$ (I) & $R_F$ (I) & $R_T$ (II) & $R_F$ (II)\\
        \hline
        SPECint & 0.93 & 0.24 & 0.92 & 0.25 & 0.26 & 0.45 & 0.22 & 0.48 & 0.14 & 0.65 & 0.13 & 0.66 & 0.74 & 0.27 & 0.75 & 0.27 \\
        SPECfp & 0.91 & 0.13 & 0.87 & 0.28 & 0.49 & 0.44 & 0.29 & 0.53 & 0.40 & 0.51 & 0.19 & 0.67 & 0.89 & 0.16 & 0.71 & 0.28 \\
        nginx & 0.92 & 0.03 & 0.91 & 0.19 & 0.68 & 0.30 & 0.35 & 0.37 & 0.47 & 0.43 & 0.24 & 0.72 & 0.89 & 0.03 & 0.67 & 0.12 \\
        postgresql & 0.80 & 0.02 & 0.75 & 0.12 & 0.45 & 0.53 & 0.25 & 0.52 & 0.28 & 0.66 & 0.23 & 0.76 & 0.74 & 0.11 & 0.42 & 0.32 \\
        trafficserver & 0.93 & 0.22 & 0.93 & 0.22 & 0.31 & 0.39 & 0.29 & 0.40 & 0.10 & 0.51 & 0.11 & 0.52 & 0.48 & 0.22 & 0.48 & 0.22 \\
        tor & 0.96 & 0.12 & 0.64 & 0.29 & 0.70 & 0.26 & 0.18 & 0.51 & 0.49 & 0.32 & 0.14 & 0.76 & 0.75 & 0.10 & 0.31 & 0.22 \\
        node & 0.99 & 0.05 & 0.95 & 0.24 & 0.74 & 0.17 & 0.31 & 0.38 & 0.64 & 0.22 & 0.28 & 0.51 & 0.92 & 0.16 & 0.69 & 0.38 \\
        \hline
    \end{tabular}
    }
\end{table*}

\begin{table*}[]
    \caption{Mean $CTR$ comparison results of our 4 CFI policies ({\it TypeArmor}, {\it IFCC}, {\it MCFI} and {\it $\tau$cfi})}\label{table:spec:ctr_short}
    \centering
    \resizebox{1.4\columnwidth}{!}{
    \begin{tabular}{|c|||r|r|r||r|r|r||r|r|r||r|r|r|}
         \hline
         \multicolumn{1}{|c|||}{} & \multicolumn{3}{|c||}{TypeArmor}
         & \multicolumn{3}{|c||}{IFCC} & \multicolumn{3}{|c||}{MCFI}
         & \multicolumn{3}{|c|}{$\tau$CFI}\\
         \hline
        Benchmark & Source & Bin-I & Bin-II & Source & Bin-I & Bin-II & Source & Bin-I & Bin-II & Source & Bin-I & Bin-II\\
        \hline
        SPECint & 3327.33 & 3966.70 & 3967.50 & 1608.22 & 606.83 & 591.66 & 1296.44 & 356.30 & 370.67 & 2105.19 & 2092.90 & 2088.97 \\
        SPECfp & 308.05 & 307.51 & 333.59 & 101.71 & 78.05 & 52.72 & 86.55 & 48.11 & 28.00 & 167.72 & 155.00 & 120.70 \\
        nginx & 506.06 & 487.85 & 570.47 & 277.97 & 201.50 & 153.71 & 130.45 & 69.33 & 142.92 & 366.34 & 357.17 & 366.55 \\
        postgresql & 6637.11 & 5337.80 & 5515.54 & 1825.17 & 1233.95 & 1009.84 & 997.45 & 720.76 & 932.53 & 2415.92 & 2372.24 & 1585.11 \\
        trafficserver & 3049.97 & 3882.86 & 3905.23 & 1866.46 & 809.56 & 754.68 & 1699.28 & 229.41 & 250.19 & 1622.12 & 990.37 & 991.23 \\
        tor & 2896.82 & 3123.42 & 2578.18 & 923.81 & 730.51 & 365.42 & 470.58 & 385.08 & 330.14 & 1610.27 & 1231.44 & 786.90 \\
        node & 70251.10 & 73847.82 & 89280.00 & 25418.30 & 23934.50 & 10240.00 & 17394.30 & 16165.50 & 8000.88 & 37759.30 & 37679.10 & 40666.90 \\
        \hline
    \end{tabular}
    }
\end{table*}

\begin{figure*}[h!]
\centering
\subfloat{
\includegraphics[width=.5\textwidth]{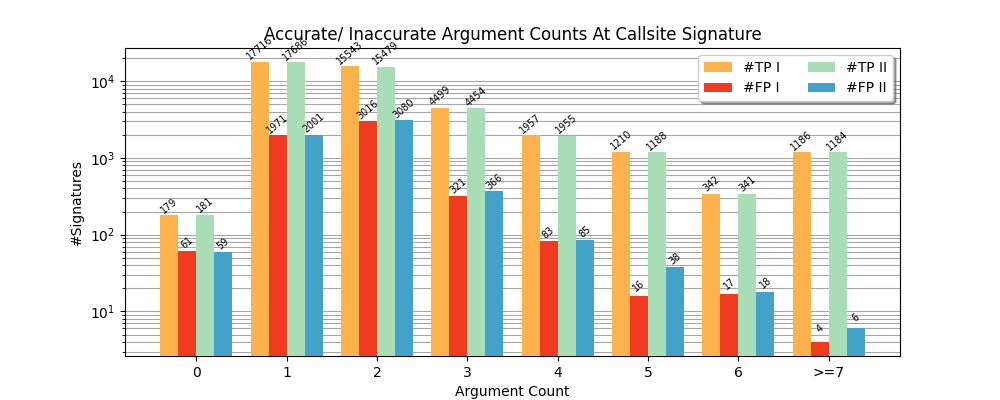}
}
\subfloat{
\includegraphics[width=.5\textwidth]{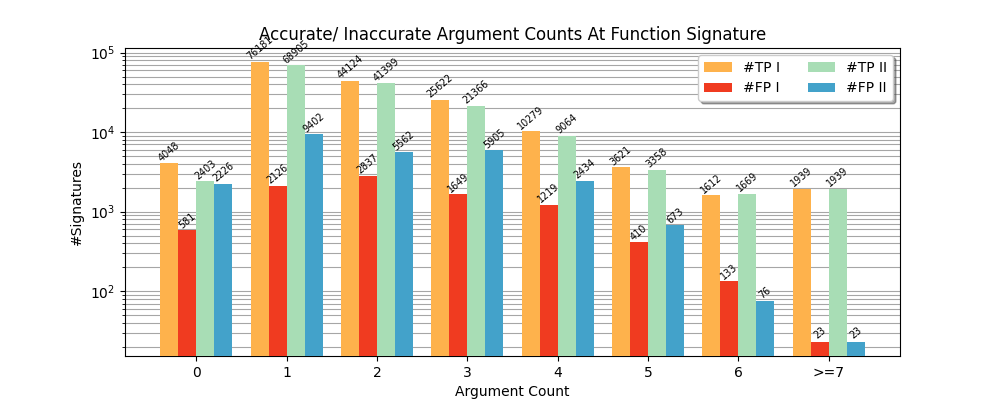}
}

\hspace*{1cm}(a) \hspace*{8.5cm} (b)
\caption{Number of True Positive and False Positive Signatures (call-site and Function) According to Argument Count recuperated over all benchmarks in Settings {\bf I} and {\bf II}}
\label{fig:argcount}
\end{figure*}

\begin{figure*}[h!]
\centering
\subfloat{
\includegraphics[width=.5\textwidth]{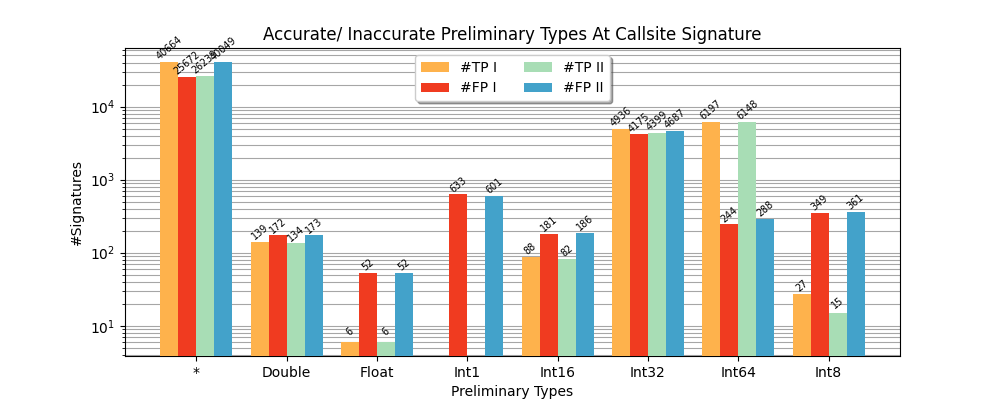}
}
\subfloat{
\includegraphics[width=.5\textwidth]{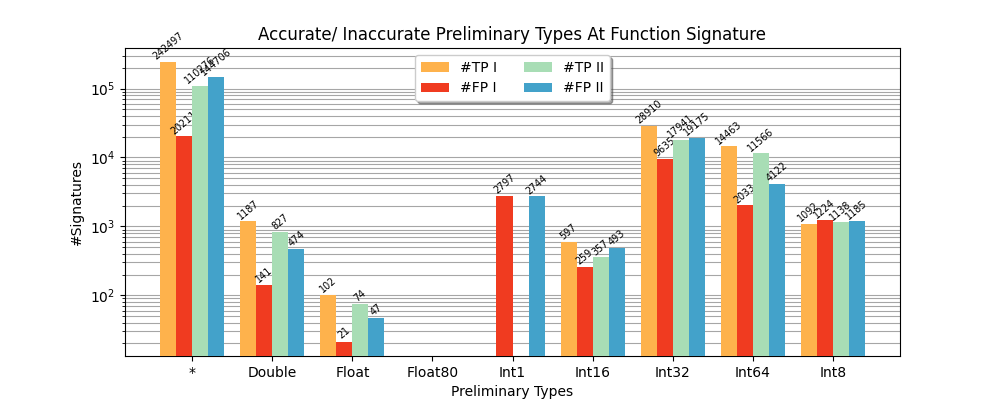}
}

\hspace*{1cm}(a) \hspace*{8.5cm} (b)
\caption{Number of True Positive and False Positive Signatures (call-site and Function) According to Preliminary Types recuperated over all benchmarks in Settings {\bf I} and {\bf II}}
\label{fig:argtypes}
\end{figure*}

\begin{figure*}[h!]
\centering
\subfloat{
\includegraphics[width=.5\textwidth]{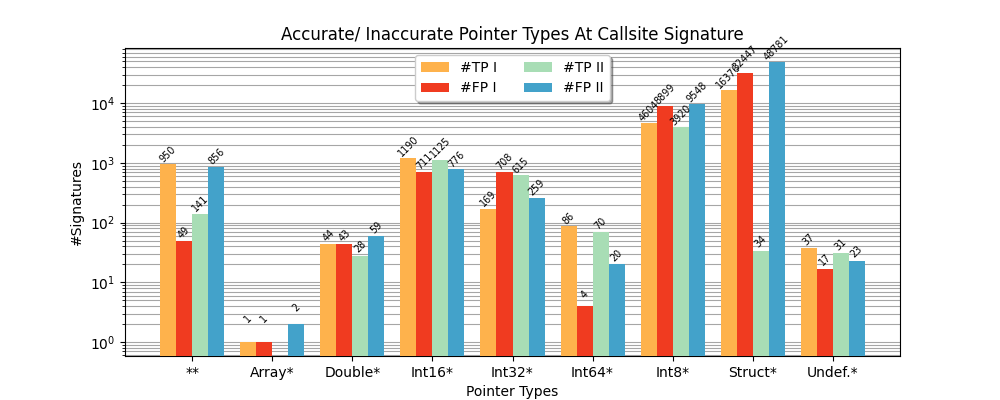}
}
\subfloat{
\includegraphics[width=.5\textwidth]{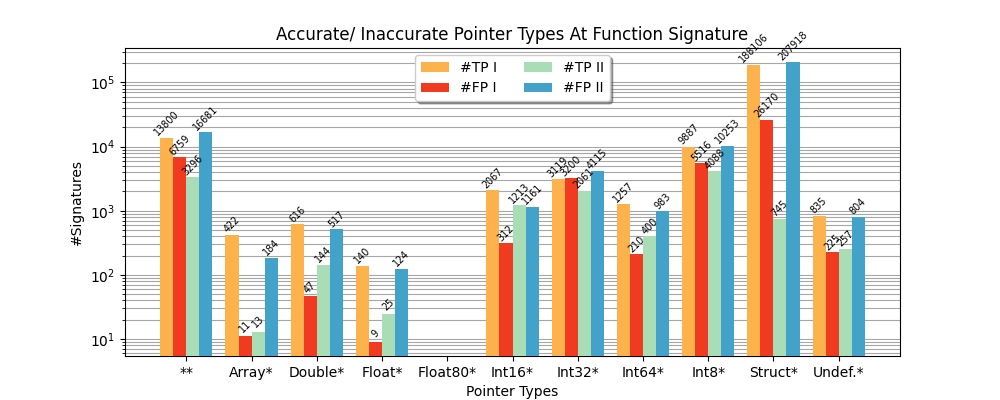}
}

\hspace*{1cm}(a) \hspace*{8.5cm} (b)
\caption{Number of True Positive and False Positive Signatures (call-site and Function) According to Pointer Types  recuperated over all benchmarks in Settings {\bf I} and {\bf II}}
\label{fig:pointertypes}
\end{figure*}

\begin{figure*}[h!]
\centering
\subfloat{
\includegraphics[width=.5\textwidth]{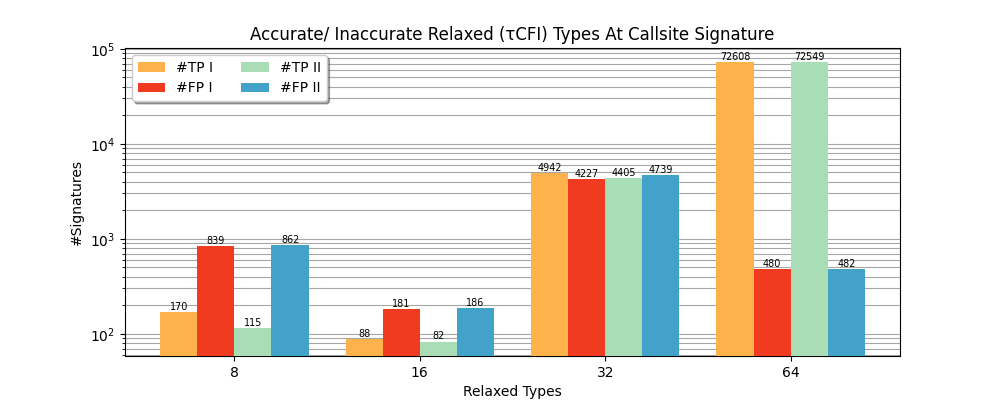}
}
\subfloat{
\includegraphics[width=.5\textwidth]{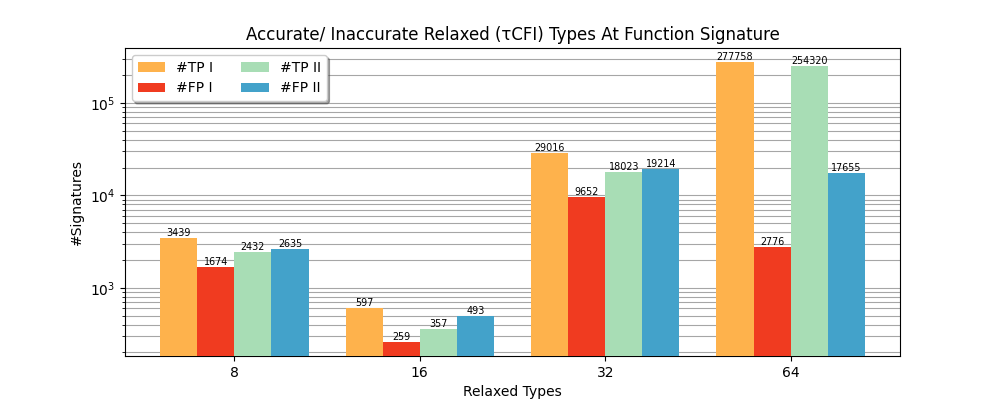}
}

\hspace*{1cm}(a) \hspace*{8.5cm} (b)
\caption{Number of True Positive and False Positive Signatures (call-site and Function) According to Relaxed Types Used in $\tau$CFI policy recuperated over all benchmarks in Settings {\bf I} and {\bf II}}
\label{fig:relaxedtypes}
\end{figure*}


\begin{table}[t]
\caption{Relaxed Return Types}\label{table:relaxed:returns}
\centering
\resizebox{1.0\columnwidth}{!}{
\begin{subtable}[]{0.7\textwidth}
\centering
\begin{tabular}{c|||c|c||c|c||c|c||c|c||c|c}
 \hline
 \multicolumn{1}{c}{} &\multicolumn{10}{c}{Call-site Signature Type} \\
 \hline
 &\#\{0\} T& \#\{0\} F& \#\{8\} T& \#\{8\} F& \#\{16\} T& \#\{16\} F& \#\{32\} T& \#\{32\} F& \#\{64\} T& \#\{64\} F\\
 \hline
 I & 22733 & 1014 & 1557 & 1357 & 538 & 155 & 2579 & 2910 & 14675 & 603\\
 \hline
 II & 21537 & 2211 & 1454 & 1461 & 214 & 479 & 1307 & 4196 & 14636 & 647 \\
 \hline
\end{tabular}
\end{subtable}
}

\resizebox{1.0\columnwidth}{!}{
\begin{subtable}[]{0.7\textwidth}
\centering
\begin{tabular}{c|||c|c||c|c||c|c||c|c||c|c}
 \hline
 \multicolumn{1}{c}{} &\multicolumn{10}{c}{Function Signature Type} \\
 \hline
 &\#\{0\} T& \#\{0\} F& \#\{8\} T& \#\{8\} F& \#\{16\} T& \#\{16\} F& \#\{32\} T& \#\{32\} F& \#\{64\} T& \#\{64\} F\\
 \hline
 I & 71243 & 7624 & 9617 & 1148 & 227 & 376 & 7480 & 9249 & 68575 & 865\\
 \hline
 II & 21012 & 57869 & 6083 & 4683 & 273 & 331 & 1087 & 15648 & 68191 & 1254 \\
 \hline
\end{tabular}
\end{subtable}
}
\end{table}

Thus, we observe that all the binary-level CFI policies modelled in this work show high levels of inaccuracy.
This inaccuracy will be manifested by the CFI policies allowing incorrect control flow transfers while tagging correct control flow transfers as erroneous at run-time.
The limitations in binary-level CFI models are caused by the imprecision in the extracted program analysis information from binaries by the SRE tools. 
Therefore, we next present and compare the accuracy of the relevant program analysis information collected by advanced SRE tools (IDA Pro, in this case).
These results help us understand the impact of imprecision in program analysis information on the CFI models that employ the data.

The \emph{TypeArmor} CFI policy needs information regarding function and call-site argument counts ({\it arity}) and the return type of each function ({\tt void} or {\tt non-void}). 
Figure~\ref{fig:argcount} and Table~\ref{table:typearmor:returns} display the accuracy of this information collected by our SRE tool, IDA Pro, compared to the ground-truth information collected by the LLVM compiler.

Figures~\ref{fig:argcount}(a) and \ref{fig:argcount}(b) categorize each call-site and function, respectively, by the number of \emph{actual} arguments in the ground-truth.
For each category of arguments plotted on the X-axis, the first pair of bars in Figure~\ref{fig:argcount}(a) shows the number of correctly and incorrectly reconstructed argument counts at each \emph{call-site} for binary configuration {\bf I}.
The next pair of bars for each category of arguments is a similar plot for binaries in configuration {\bf II}.
The results are shown on a logarithmic scale.
We find that, overall, call-site argument counts are correctly detected in 89\% of the cases in settings {\bf I} and 88\% of the cases in settings {\bf II}.

Figure~\ref{fig:argcount}(b) is a similar graph that plots the number of correctly and incorrectly reconstructed function parameter counts for binary configuration {\bf I} (first pair of bars for each category of parameter counts) and binary configuration {\bf II} (second pair of bars for each parameter counts category), respectively.
We observe that 95\% of function parameters are accurately detected in settings {\bf I} and 85\% parameters are accurately detected in settings {\bf II}, on average.
\begin{mybox}
\textit{\textbf{Observation 5.} 
Overall, we find that state-of-the-art SRE tools can accurately detect the number of call-site and function argument counts in most cases.
Also interesting is the observation that the lack of symbol information (in config. {\bf II}) does not significantly affect the accuracy of argument count detection.
This high accuracy is reflected in the relatively high $RelativeCTR_T$ and low $RelativeCTR_F$ numbers for most benchmark suites in Table~\ref{table:spec-opt-stripped:rctr_short}.
}
\end{mybox}

\begin{table}[]
\caption{Void/Non-Void Return Types}
\label{table:typearmor:returns}
\centering
\resizebox{0.6\columnwidth}{!}{
\begin{subtable}[]{0.4\textwidth}
\begin{tabular}{c|||c|c||c|c}
 \hline
 \multicolumn{1}{c}{} & \multicolumn{4}{c}{Call-site Signature Type} \\
 \hline
 & \#Void T& \#Void F& \#Non-Void T& \#Non-Void F\\
 \hline
 I & 22733 & 1014 & 21955 & 2419 \\
 \hline
 II &21537 & 2211 & 21999 & 2395\\
 \hline
\end{tabular}
\end{subtable}
}
\resizebox{0.6\columnwidth}{!}{
\begin{subtable}[]{0.4\textwidth}
\begin{tabular}{c|||c|c||c|c}
 \hline
 \multicolumn{1}{c}{} &\multicolumn{4}{c}{Function Signature Type} \\
 \hline
 &\#Void T& \#Void F& \#Non-Void T& \#Non-Void F\\
 \hline
 I & 71243 & 7624 & 96844 & 693\\
 \hline
 II & 21003 & 57864 & 96777 & 760 \\
 \hline
\end{tabular}
\end{subtable}
}
\end{table}

Table~\ref{table:typearmor:returns} shows the number of {\tt void} and {\tt non-void} call-sites and functions that are (in)correctly identified by the IDA Pro SRE tool.
The columns labeled {\it `Void T'} and {\it `Non-Void T'} in Table~\ref{table:typearmor:returns} denote the number of correctly identified call-sites and functions with {\tt void}  and {\tt non-void} as the \emph{actual} return type, respectively.
Likewise, columns labeled {\it `Void F'} and {\it `Non-Void F'} denote the number of incorrectly identified call-sites and functions with the {\tt void} and {\tt non-void} \emph{actual} return type, respectively.
Thus, we find that the advanced SRE tools available today can correctly detect over 95\% and 89\% of call-site instances, and 90\% and 99\% of function instances with the {\tt void} and {\tt non-void} return type, respectively, for binaries in configuration {\bf I}.
While the detection of call-site instances and {\tt non-void} function instances is similarly accurate even for binaries in configuration {\bf II}, the accuracy of {\tt void} function detection degrades to around 27\%.
Many non-returning functions are portrayed as returning due to incorrect static analysis for configuration {\bf II} binaries.
Thus, we make following interesting observation.
\begin{mybox}
\textit{\textbf{Observation 6.} Detection of {\tt void} and {\tt non-void} function and call-site return type inference is fairly accurate except for detection of {\tt void} functions when debug symbol information is missing.
}
\end{mybox}


The {\it IFCC} and {\it MCFI} CFI policies consider mature types to apply policy constraints. 
The only difference between these two policies is that while {\it MCFI} distinguishes between multi-level pointers, like {\tt void*} and {\tt int*}, the {\it IFCC} policy only considers single-level pointers.
Figures~\ref{fig:argtypes} and~\ref{fig:pointertypes} display detection accuracy of all the types at function and call-site recuperated over all benchmarks. 
Each figure shows two plots, the first compares the call-site signatures and the second compares function signatures in the binary {\bf I} and {\bf II} configurations with the ground truth.
The X-axis in each figure categorizes the types considered.
Similar to Figure~\ref{fig:argcount}, the the first and second pairs of bars in each category plots the true positive and false positive numbers for binary configurations {\bf I} and {\bf II}, respectively.
While Figure~\ref{fig:argtypes} compares preliminary types, Figure~\ref{fig:pointertypes} compares the base types of pointers. 

We observe in Figure~\ref{fig:argtypes} that the accuracy of preliminary type detection at call-sites and functions is around 62\% and 89\% respectively in setting {\bf I}. 
However, the accuracy decreases significantly (around 44\% and 45\% for call-site and function argument types) when the symbols aren't available in setting {\bf II}. 
On the other hand, the detection accuracy of base pointer types is around 35\% and 84\% at call-site and call-targets respectively in setting {\bf I}, but the accuracy is extremely low i.e. around 9\% and 5\% in setting {\bf II}. 
With some manual analysis with the {\tt Nginx} benchmark, we found that the mischaracterization of the {\tt struct*} type as {\tt int64} or {\tt int64*} by the binary analysis tool is one important reason for the high error rate.
Thus, we make following interesting observation.
\begin{mybox}
\textit{\textbf{Observation 7.} The poor preliminary and pointer type detection by the SRE tools, especially with binary configuration {\bf II}, results in the high error rates witnessed in the MCFI and IFCC CFI policies at the binary level.
}
\end{mybox}

The $\tau$CFI policy considers {\it relaxed} types in range {\tt \{0,8,16,32,64\}} according to the width of registers used to store parameters.
Figure~\ref{fig:relaxedtypes} displays the true and false positive counts of relaxed types. 
The accuracy in relaxed type detection by the binary analysis tool is around 94\% and 96\% for call-site and function types in setting {\bf I}, and 93\% and 87\% for call-site and function types in setting {\bf II}, respectively. 
As the $\tau$CFI policy considers return types as well, we also show the correctly and incorrectly detected types in Table~\ref{table:relaxed:returns}. Observe that the accuracy decays by a notable amount in setting {\bf II}, especially when the size is {\tt 32} bits. 
Overall, we find that, 
\begin{mybox}
\textit{\textbf{Observation 8.}
Binary level analysis tools achieve reasonable accuracy in relaxed type detection, especially in binary configuration {\bf I}, which is responsible for the relatively lower error rates with the $\tau$CFI policy, compared with MCFI and IFCC CFI policies.
}
\end{mybox}

Finally, we notice that SRE tools like IDA Pro do not take full advantage of \textit{Dwarf} symbols during their analysis. 
One reason for not using this information, even when available, may be that the {\it Dwarf} information may not be trustworthy. 
We hope that our observations in this section assist binary analysis tools and CFI evaluators in offering improved CFI results with binary level policies.

\section{Limitations and Future Work}
\label{future}
There are several avenues for future work.
Firstly, the definitions of true/false positives/negatives used in this work only measure how accurately the binary CFI policy follows its designed algorithm to categorize the set of target functions at each call-site into \emph{reachable} and \emph{unreachable} sets.
Accordingly, the overestimation of the set of reachable targets by most (source-level) CFI policies as well as the non-uniform distribution of gadgets in the program implies that the \emph{false positives} and \emph{false negatives} reported for the evaluated binary-level CFI policies, although \emph{incorrect} according to the algorithm, may not actually cause a legal program execution to necessarily fail or increase the program vulnerability at run-time. 
We will attempt to better understand the impact of the \emph{false}-ness in binary-level CFI policies in future work.


Secondly, 
although the CFI policy comparison framework developed in this work is highly modular, we currently only integrate a single binary analysis module (the IDA Pro SRE tool) to extract binary-level program analysis information.
In the future, we intend to leverage analysis information from different advanced binary-level SRE tools to understand their impact on CFI accuracy. In addition to that, it is also interesting to leverage recent type inference techniques~\cite{noonan2016,Chua2017,Ligeng2020,Kexin2021,Zhang2021,Chen2022} and investigate the affect on accuracy over various type-based CFI policies.

Lastly, we only study type-based CFI policies in this work.
In the future we will augment and improve our current target set analysis by using an advanced type propagation~\cite{lu2019, bauer2022} and pointer analysis systems~\cite{fan2017} and study their impact on binary-level CFI techniques.
We will also study other CFI policies that can manage advanced constructs, like polymorphism.


\section{Conclusion}
\label{conclusion}
Our goal in this work was to explore and quantify the precision of binary-level CFI techniques, and understand how that precision is impacted by the inaccuracies in the program analysis information recovered by even the most modern SRE tools.
We developed a comprehensive infrastructure, a thorough mechanism, and new metrics to achieve this goal.
Our modular framework can model and evaluate different binary-level type-based CFI policies by comparing their outcomes with their source-based counterparts.
We demonstrated our framework and reported results for four binary-level CFI policies using program information from the IDA Pro binary analysis tool, compared to a source-level ground truth implemented by a new LTO pass in the LLVM compiler.

The results achieved by our novel mechanism and metrics highlight the unresolved challenges for modern SRE tools in correctly extracting the relevant program information, and their potentially staggering impact on the precision of binary-level CFI techniques that use such data.
Given the current state of binary-level tools, we find that the more advanced CFI techniques that employ finer-grained type information and offer impressive protection from control-flow hijacking attacks at the \emph{source} level, will achieve extremely poor performance if applied at the binary level with shockingly high number of false positives and false negatives.
We expect our work will help researchers and engineers built more advanced binary-level analysis algorithms and CFI techniques, and more accurately measure their precision and outcomes.


\bibliographystyle{ACM-Reference-Format}
\bibliography{acsac}


\begin{thebibliography}{43}


\ifx \showCODEN    \undefined \def \showCODEN     #1{\unskip}     \fi
\ifx \showDOI      \undefined \def \showDOI       #1{#1}\fi
\ifx \showISBNx    \undefined \def \showISBNx     #1{\unskip}     \fi
\ifx \showISBNxiii \undefined \def \showISBNxiii  #1{\unskip}     \fi
\ifx \showISSN     \undefined \def \showISSN      #1{\unskip}     \fi
\ifx \showLCCN     \undefined \def \showLCCN      #1{\unskip}     \fi
\ifx \shownote     \undefined \def \shownote      #1{#1}          \fi
\ifx \showarticletitle \undefined \def \showarticletitle #1{#1}   \fi
\ifx \showURL      \undefined \def \showURL       {\relax}        \fi
\providecommand\bibfield[2]{#2}
\providecommand\bibinfo[2]{#2}
\providecommand\natexlab[1]{#1}
\providecommand\showeprint[2][]{arXiv:#2}

\bibitem[Abadi et~al\mbox{.}(2005)]%
        {abadi2005}
\bibfield{author}{\bibinfo{person}{Mart\'{\i}n Abadi}, \bibinfo{person}{Mihai
  Budiu}, \bibinfo{person}{\'{U}lfar Erlingsson}, {and} \bibinfo{person}{Jay
  Ligatti}.} \bibinfo{year}{2005}\natexlab{}.
\newblock \showarticletitle{Control-Flow Integrity}. In
  \bibinfo{booktitle}{\emph{Proceedings of the 12th ACM Conference on Computer
  and Communications Security}} (Alexandria, VA, USA)
  \emph{(\bibinfo{series}{CCS '05})}. \bibinfo{publisher}{Association for
  Computing Machinery}, \bibinfo{address}{New York, NY, USA},
  \bibinfo{pages}{340–353}.
\newblock


\bibitem[Andriesse et~al\mbox{.}(2016)]%
        {Andriesse2016}
\bibfield{author}{\bibinfo{person}{Dennis Andriesse}, \bibinfo{person}{Xi
  Chen}, \bibinfo{person}{Victor van~der Veen}, \bibinfo{person}{Asia
  Slowinska}, {and} \bibinfo{person}{Herbert Bos}.}
  \bibinfo{year}{2016}\natexlab{}.
\newblock \showarticletitle{An {In-Depth} Analysis of Disassembly on
  {Full-Scale} x86/x64 Binaries}. In \bibinfo{booktitle}{\emph{25th USENIX
  Security Symposium (USENIX Security 16)}}. \bibinfo{publisher}{USENIX
  Association}, \bibinfo{address}{Austin, TX}, \bibinfo{pages}{583--600}.
\newblock
\showISBNx{978-1-931971-32-4}
\urldef\tempurl%
\url{https://www.usenix.org/conference/usenixsecurity16/technical-sessions/presentation/andriesse}
\showURL{%
\tempurl}


\bibitem[Bauer et~al\mbox{.}(2022)]%
        {bauer2022}
\bibfield{author}{\bibinfo{person}{Markus Bauer}, \bibinfo{person}{Ilya
  Grishchenko}, {and} \bibinfo{person}{Christian Rossow}.}
  \bibinfo{year}{2022}\natexlab{}.
\newblock \showarticletitle{TyPro: Forward CFI for C-Style Indirect Function
  Calls Using Type Propagation}. In \bibinfo{booktitle}{\emph{Proceedings of
  the 38th Annual Computer Security Applications Conference}} (Austin, TX, USA)
  \emph{(\bibinfo{series}{ACSAC '22})}. \bibinfo{publisher}{Association for
  Computing Machinery}, \bibinfo{address}{New York, NY, USA},
  \bibinfo{pages}{346–360}.
\newblock
\showISBNx{9781450397599}
\urldef\tempurl%
\url{https://doi.org/10.1145/3564625.3564627}
\showDOI{\tempurl}


\bibitem[Burow et~al\mbox{.}(2017)]%
        {Burow2017}
\bibfield{author}{\bibinfo{person}{Nathan Burow}, \bibinfo{person}{Scott~A.
  Carr}, \bibinfo{person}{Joseph Nash}, \bibinfo{person}{Per Larsen},
  \bibinfo{person}{Michael Franz}, \bibinfo{person}{Stefan Brunthaler}, {and}
  \bibinfo{person}{Mathias Payer}.} \bibinfo{year}{2017}\natexlab{}.
\newblock \showarticletitle{Control-Flow Integrity: Precision, Security, and
  Performance}.
\newblock \bibinfo{journal}{\emph{ACM Comput. Surv.}} \bibinfo{volume}{50},
  \bibinfo{number}{1}, Article \bibinfo{articleno}{16} (\bibinfo{date}{apr}
  \bibinfo{year}{2017}), \bibinfo{numpages}{33}~pages.
\newblock
\showISSN{0360-0300}
\urldef\tempurl%
\url{https://doi.org/10.1145/3054924}
\showDOI{\tempurl}


\bibitem[Caballero and Lin(2016)]%
        {caballero2016}
\bibfield{author}{\bibinfo{person}{Juan Caballero} {and}
  \bibinfo{person}{Zhiqiang Lin}.} \bibinfo{year}{2016}\natexlab{}.
\newblock \showarticletitle{Type Inference on Executables}.
\newblock \bibinfo{journal}{\emph{ACM Comput. Surv.}} \bibinfo{volume}{48},
  \bibinfo{number}{4}, Article \bibinfo{articleno}{65} (\bibinfo{date}{may}
  \bibinfo{year}{2016}), \bibinfo{numpages}{35}~pages.
\newblock
\showISSN{0360-0300}
\urldef\tempurl%
\url{https://doi.org/10.1145/2896499}
\showDOI{\tempurl}


\bibitem[Checkoway et~al\mbox{.}(2010)]%
        {checkoway2010}
\bibfield{author}{\bibinfo{person}{Stephen Checkoway}, \bibinfo{person}{Lucas
  Davi}, \bibinfo{person}{Alexandra Dmitrienko}, \bibinfo{person}{Ahmad-Reza
  Sadeghi}, \bibinfo{person}{Hovav Shacham}, {and} \bibinfo{person}{Marcel
  Winandy}.} \bibinfo{year}{2010}\natexlab{}.
\newblock \showarticletitle{Return-Oriented Programming without Returns}. In
  \bibinfo{booktitle}{\emph{Proceedings of the 17th ACM Conference on Computer
  and Communications Security}} (Chicago, Illinois, USA)
  \emph{(\bibinfo{series}{CCS '10})}. \bibinfo{publisher}{Association for
  Computing Machinery}, \bibinfo{address}{New York, NY, USA},
  \bibinfo{pages}{559–572}.
\newblock
\showISBNx{9781450302456}
\urldef\tempurl%
\url{https://doi.org/10.1145/1866307.1866370}
\showDOI{\tempurl}


\bibitem[Chen et~al\mbox{.}(2020)]%
        {Ligeng2020}
\bibfield{author}{\bibinfo{person}{Ligeng Chen}, \bibinfo{person}{Zhongling
  He}, {and} \bibinfo{person}{Bing Mao}.} \bibinfo{year}{2020}\natexlab{}.
\newblock \showarticletitle{CATI: Context-Assisted Type Inference from Stripped
  Binaries}. In \bibinfo{booktitle}{\emph{2020 50th Annual IEEE/IFIP
  International Conference on Dependable Systems and Networks (DSN)}}.
  \bibinfo{pages}{88--98}.
\newblock
\urldef\tempurl%
\url{https://doi.org/10.1109/DSN48063.2020.00028}
\showDOI{\tempurl}


\bibitem[Chen et~al\mbox{.}(2022)]%
        {Chen2022}
\bibfield{author}{\bibinfo{person}{Qibin Chen}, \bibinfo{person}{Jeremy
  Lacomis}, \bibinfo{person}{Edward~J. Schwartz}, \bibinfo{person}{Claire~Le
  Goues}, \bibinfo{person}{Graham Neubig}, {and} \bibinfo{person}{Bogdan
  Vasilescu}.} \bibinfo{year}{2022}\natexlab{}.
\newblock \showarticletitle{Augmenting Decompiler Output with Learned Variable
  Names and Types}. In \bibinfo{booktitle}{\emph{31st USENIX Security Symposium
  (USENIX Security 22)}}. \bibinfo{publisher}{USENIX Association},
  \bibinfo{address}{Boston, MA}, \bibinfo{pages}{4327--4343}.
\newblock
\showISBNx{978-1-939133-31-1}
\urldef\tempurl%
\url{https://www.usenix.org/conference/usenixsecurity22/presentation/chen-qibin}
\showURL{%
\tempurl}


\bibitem[Chua et~al\mbox{.}(2017)]%
        {Chua2017}
\bibfield{author}{\bibinfo{person}{Zheng~Leong Chua}, \bibinfo{person}{Shiqi
  Shen}, \bibinfo{person}{Prateek Saxena}, {and} \bibinfo{person}{Zhenkai
  Liang}.} \bibinfo{year}{2017}\natexlab{}.
\newblock \showarticletitle{Neural Nets Can Learn Function Type Signatures From
  Binaries}. In \bibinfo{booktitle}{\emph{26th USENIX Security Symposium
  (USENIX Security 17)}}. \bibinfo{publisher}{USENIX Association},
  \bibinfo{address}{Vancouver, BC}, \bibinfo{pages}{99--116}.
\newblock
\showISBNx{978-1-931971-40-9}
\urldef\tempurl%
\url{https://www.usenix.org/conference/usenixsecurity17/technical-sessions/presentation/chua}
\showURL{%
\tempurl}


\bibitem[Crane et~al\mbox{.}(2015)]%
        {Crane2015}
\bibfield{author}{\bibinfo{person}{Stephen~J. Crane}, \bibinfo{person}{Stijn
  Volckaert}, \bibinfo{person}{Felix Schuster}, \bibinfo{person}{Christopher
  Liebchen}, \bibinfo{person}{Per Larsen}, \bibinfo{person}{Lucas Davi},
  \bibinfo{person}{Ahmad-Reza Sadeghi}, \bibinfo{person}{Thorsten Holz},
  \bibinfo{person}{Bjorn De~Sutter}, {and} \bibinfo{person}{Michael Franz}.}
  \bibinfo{year}{2015}\natexlab{}.
\newblock \showarticletitle{It's a TRaP: Table Randomization and Protection
  against Function-Reuse Attacks}. In \bibinfo{booktitle}{\emph{Proceedings of
  the 22nd ACM SIGSAC Conference on Computer and Communications Security}}
  (Denver, Colorado, USA) \emph{(\bibinfo{series}{CCS '15})}.
  \bibinfo{publisher}{Association for Computing Machinery},
  \bibinfo{address}{New York, NY, USA}, \bibinfo{pages}{243–255}.
\newblock
\showISBNx{9781450338325}
\urldef\tempurl%
\url{https://doi.org/10.1145/2810103.2813682}
\showDOI{\tempurl}


\bibitem[Designer(1997)]%
        {designer1997}
\bibfield{author}{\bibinfo{person}{Solar Designer}.}
  \bibinfo{year}{1997}\natexlab{}.
\newblock \showarticletitle{Getting around non-executable stack (and fix)}.
\newblock
  \bibinfo{journal}{\emph{"http://ouah.bsdjeunz.org/solarretlibc.html"}}
  (\bibinfo{year}{1997}).
\newblock


\bibitem[Fan et~al\mbox{.}(2017)]%
        {fan2017}
\bibfield{author}{\bibinfo{person}{Xiaokang Fan}, \bibinfo{person}{Yulei Sui},
  \bibinfo{person}{Xiangke Liao}, {and} \bibinfo{person}{Jingling Xue}.}
  \bibinfo{year}{2017}\natexlab{}.
\newblock \showarticletitle{Boosting the Precision of Virtual Call Integrity
  Protection with Partial Pointer Analysis for C++}. In
  \bibinfo{booktitle}{\emph{Proceedings of the 26th ACM SIGSOFT International
  Symposium on Software Testing and Analysis}} (Santa Barbara, CA, USA)
  \emph{(\bibinfo{series}{ISSTA 2017})}. \bibinfo{publisher}{Association for
  Computing Machinery}, \bibinfo{address}{New York, NY, USA},
  \bibinfo{pages}{329–340}.
\newblock
\showISBNx{9781450350761}
\urldef\tempurl%
\url{https://doi.org/10.1145/3092703.3092729}
\showDOI{\tempurl}


\bibitem[Farkhani et~al\mbox{.}(2018)]%
        {Farkhani2018}
\bibfield{author}{\bibinfo{person}{Reza~Mirzazade Farkhani},
  \bibinfo{person}{Saman Jafari}, \bibinfo{person}{Sajjad Arshad},
  \bibinfo{person}{William Robertson}, \bibinfo{person}{Engin Kirda}, {and}
  \bibinfo{person}{Hamed Okhravi}.} \bibinfo{year}{2018}\natexlab{}.
\newblock \showarticletitle{On the Effectiveness of Type-Based Control Flow
  Integrity}. In \bibinfo{booktitle}{\emph{Proceedings of the 34th Annual
  Computer Security Applications Conference}} (San Juan, PR, USA)
  \emph{(\bibinfo{series}{ACSAC '18})}. \bibinfo{publisher}{Association for
  Computing Machinery}, \bibinfo{address}{New York, NY, USA},
  \bibinfo{pages}{28–39}.
\newblock
\showISBNx{9781450365697}
\urldef\tempurl%
\url{https://doi.org/10.1145/3274694.3274739}
\showDOI{\tempurl}


\bibitem[Frassetto et~al\mbox{.}(2022)]%
        {Frassetto2022}
\bibfield{author}{\bibinfo{person}{Tommaso Frassetto}, \bibinfo{person}{Patrick
  Jauernig}, \bibinfo{person}{David Koisser}, {and} \bibinfo{person}{Ahmad-Reza
  Sadeghi}.} \bibinfo{year}{2022}\natexlab{}.
\newblock \showarticletitle{CFInsight: A Comprehensive Metric for CFI
  Policies}. In \bibinfo{booktitle}{\emph{NDSS}}.
\newblock
\urldef\tempurl%
\url{https://doi.org/10.14722/ndss.2022.23165}
\showDOI{\tempurl}


\bibitem[gcc~fcf protection(2022)]%
        {gcccfi2023}
\bibfield{author}{\bibinfo{person}{gcc~fcf protection}.}
  \bibinfo{year}{2022}\natexlab{}.
\newblock
  \showarticletitle{https://gcc.gnu.org/onlinedocs/gcc/Instrumentation-Options.html}.
  In \bibinfo{booktitle}{\emph{Clang 16.0.0git documentation}}.
\newblock


\bibitem[Ge et~al\mbox{.}(2016)]%
        {aia2016}
\bibfield{author}{\bibinfo{person}{Xinyang Ge}, \bibinfo{person}{Nirupama
  Talele}, \bibinfo{person}{Mathias Payer}, {and} \bibinfo{person}{Trent
  Jaeger}.} \bibinfo{year}{2016}\natexlab{}.
\newblock \showarticletitle{Fine-Grained Control-Flow Integrity for Kernel
  Software}. In \bibinfo{booktitle}{\emph{2016 IEEE European Symposium on
  Security and Privacy (EuroS\&P)}}. \bibinfo{pages}{179--194}.
\newblock
\urldef\tempurl%
\url{https://doi.org/10.1109/EuroSP.2016.24}
\showDOI{\tempurl}


\bibitem[IDA~Pro(2022)]%
        {idapro2022}
\bibfield{author}{\bibinfo{person}{hexrays IDA~Pro}.}
  \bibinfo{year}{2022}\natexlab{}.
\newblock \showarticletitle{https://hex-rays.com/ida-pro/}. In
  \bibinfo{booktitle}{\emph{Interactive Disassembler (IDA)}}.
\newblock


\bibitem[Lan et~al\mbox{.}(2015)]%
        {Lan2015}
\bibfield{author}{\bibinfo{person}{Bingchen Lan}, \bibinfo{person}{Yan Li},
  \bibinfo{person}{Hao Sun}, \bibinfo{person}{Chao Su}, \bibinfo{person}{Yao
  Liu}, {and} \bibinfo{person}{Qingkai Zeng}.} \bibinfo{year}{2015}\natexlab{}.
\newblock \showarticletitle{Loop-Oriented Programming: A New Code Reuse Attack
  to Bypass Modern Defenses}. In \bibinfo{booktitle}{\emph{2015 IEEE
  Trustcom/BigDataSE/ISPA}}, Vol.~\bibinfo{volume}{1}.
  \bibinfo{pages}{190--197}.
\newblock
\urldef\tempurl%
\url{https://doi.org/10.1109/Trustcom.2015.374}
\showDOI{\tempurl}


\bibitem[Lettner et~al\mbox{.}(2016)]%
        {Lettner2016}
\bibfield{author}{\bibinfo{person}{Julian Lettner}, \bibinfo{person}{Benjamin
  Kollenda}, \bibinfo{person}{Andrei Homescu}, \bibinfo{person}{Per Larsen},
  \bibinfo{person}{Felix Schuster}, \bibinfo{person}{Lucas Davi},
  \bibinfo{person}{Ahmad-Reza Sadeghi}, \bibinfo{person}{Thorsten Holz}, {and}
  \bibinfo{person}{Michael Franz}.} \bibinfo{year}{2016}\natexlab{}.
\newblock \showarticletitle{{Subversive-C}: Abusing and Protecting Dynamic
  Message Dispatch}. In \bibinfo{booktitle}{\emph{2016 USENIX Annual Technical
  Conference (USENIX ATC 16)}}. \bibinfo{publisher}{USENIX Association},
  \bibinfo{address}{Denver, CO}, \bibinfo{pages}{209--221}.
\newblock
\showISBNx{978-1-931971-30-0}
\urldef\tempurl%
\url{https://www.usenix.org/conference/atc16/technical-sessions/presentation/lettner}
\showURL{%
\tempurl}


\bibitem[Li et~al\mbox{.}(2020)]%
        {cracks2020}
\bibfield{author}{\bibinfo{person}{Yuan Li}, \bibinfo{person}{Mingzhe Wang},
  \bibinfo{person}{Chao Zhang}, \bibinfo{person}{Xingman Chen},
  \bibinfo{person}{Songtao Yang}, {and} \bibinfo{person}{Ying Liu}.}
  \bibinfo{year}{2020}\natexlab{}.
\newblock \showarticletitle{Finding Cracks in Shields: On the Security of
  Control Flow Integrity Mechanisms}. In \bibinfo{booktitle}{\emph{Proceedings
  of the 2020 ACM SIGSAC Conference on Computer and Communications Security}}
  (Virtual Event, USA) \emph{(\bibinfo{series}{CCS '20})}.
  \bibinfo{publisher}{Association for Computing Machinery},
  \bibinfo{address}{New York, NY, USA}, \bibinfo{pages}{1821–1835}.
\newblock
\showISBNx{9781450370899}
\urldef\tempurl%
\url{https://doi.org/10.1145/3372297.3417867}
\showDOI{\tempurl}


\bibitem[Liu and Wang(2020)]%
        {liu2020}
\bibfield{author}{\bibinfo{person}{Zhibo Liu} {and} \bibinfo{person}{Shuai
  Wang}.} \bibinfo{year}{2020}\natexlab{}.
\newblock \showarticletitle{How Far We Have Come: Testing Decompilation
  Correctness of C Decompilers}. In \bibinfo{booktitle}{\emph{Proceedings of
  the 29th ACM SIGSOFT International Symposium on Software Testing and
  Analysis}} (Virtual Event, USA) \emph{(\bibinfo{series}{ISSTA 2020})}.
  \bibinfo{publisher}{Association for Computing Machinery},
  \bibinfo{address}{New York, NY, USA}, \bibinfo{pages}{475–487}.
\newblock
\showISBNx{9781450380089}
\urldef\tempurl%
\url{https://doi.org/10.1145/3395363.3397370}
\showDOI{\tempurl}


\bibitem[LLVM(2022)]%
        {llvmcfi2022}
\bibfield{author}{\bibinfo{person}{LLVM}.} \bibinfo{year}{2022}\natexlab{}.
\newblock
  \showarticletitle{https://clang.llvm.org/docs/ControlFlowIntegrity.html}. In
  \bibinfo{booktitle}{\emph{Clang 16.0.0git documentation}}.
\newblock


\bibitem[LLVM(2023)]%
        {llvm2023}
\bibfield{author}{\bibinfo{person}{LLVM}.} \bibinfo{year}{2023}\natexlab{}.
\newblock \showarticletitle{https://llvm.org}. In \bibinfo{booktitle}{\emph{The
  LLVM Compiler Infrastructure}}.
\newblock


\bibitem[Lu and Hu(2019)]%
        {lu2019}
\bibfield{author}{\bibinfo{person}{Kangjie Lu} {and} \bibinfo{person}{Hong
  Hu}.} \bibinfo{year}{2019}\natexlab{}.
\newblock \showarticletitle{Where Does It Go? Refining Indirect-Call Targets
  with Multi-Layer Type Analysis}. In \bibinfo{booktitle}{\emph{Proceedings of
  the 2019 ACM SIGSAC Conference on Computer and Communications Security}}
  (London, United Kingdom) \emph{(\bibinfo{series}{CCS '19})}.
  \bibinfo{publisher}{Association for Computing Machinery},
  \bibinfo{address}{New York, NY, USA}, \bibinfo{pages}{1867–1881}.
\newblock
\showISBNx{9781450367479}
\urldef\tempurl%
\url{https://doi.org/10.1145/3319535.3354244}
\showDOI{\tempurl}


\bibitem[Meng and Miller(2016)]%
        {meng2016}
\bibfield{author}{\bibinfo{person}{Xiaozhu Meng} {and}
  \bibinfo{person}{Barton~P. Miller}.} \bibinfo{year}{2016}\natexlab{}.
\newblock \showarticletitle{Binary Code is Not Easy}. In
  \bibinfo{booktitle}{\emph{Proceedings of the 25th International Symposium on
  Software Testing and Analysis}} (Saarbr\"{u}cken, Germany)
  \emph{(\bibinfo{series}{ISSTA 2016})}. \bibinfo{publisher}{Association for
  Computing Machinery}, \bibinfo{address}{New York, NY, USA},
  \bibinfo{pages}{24–35}.
\newblock
\showISBNx{9781450343909}
\urldef\tempurl%
\url{https://doi.org/10.1145/2931037.2931047}
\showDOI{\tempurl}


\bibitem[Muntean et~al\mbox{.}(2018)]%
        {Muntean2018}
\bibfield{author}{\bibinfo{person}{Paul Muntean}, \bibinfo{person}{Matthias
  Fischer}, \bibinfo{person}{Gang Tan}, \bibinfo{person}{Zhiqiang Lin},
  \bibinfo{person}{Jens Grossklags}, {and} \bibinfo{person}{Claudia Eckert}.}
  \bibinfo{year}{2018}\natexlab{}.
\newblock \showarticletitle{$\tau$CFI: Type-Assisted Control Flow Integrity for
  x86-64 Binaries}. In \bibinfo{booktitle}{\emph{Research in Attacks,
  Intrusions, and Defenses}}. \bibinfo{publisher}{Springer International
  Publishing}, \bibinfo{address}{Cham}, \bibinfo{pages}{423--444}.
\newblock
\showISBNx{978-3-030-00470-5}


\bibitem[Muntean et~al\mbox{.}(2019)]%
        {Muntean2019}
\bibfield{author}{\bibinfo{person}{Paul Muntean}, \bibinfo{person}{Matthias
  Neumayer}, \bibinfo{person}{Zhiqiang Lin}, \bibinfo{person}{Gang Tan},
  \bibinfo{person}{Jens Grossklags}, {and} \bibinfo{person}{Claudia Eckert}.}
  \bibinfo{year}{2019}\natexlab{}.
\newblock \showarticletitle{Analyzing Control Flow Integrity with LLVM-CFI}. In
  \bibinfo{booktitle}{\emph{Proceedings of the 35th Annual Computer Security
  Applications Conference}} (San Juan, Puerto Rico, USA)
  \emph{(\bibinfo{series}{ACSAC '19})}. \bibinfo{publisher}{Association for
  Computing Machinery}, \bibinfo{address}{New York, NY, USA},
  \bibinfo{pages}{584–597}.
\newblock


\bibitem[Niu and Tan(2014)]%
        {mcfi2014}
\bibfield{author}{\bibinfo{person}{Ben Niu} {and} \bibinfo{person}{Gang Tan}.}
  \bibinfo{year}{2014}\natexlab{}.
\newblock \showarticletitle{Modular Control-Flow Integrity}.
\newblock \bibinfo{journal}{\emph{SIGPLAN Not.}} \bibinfo{volume}{49},
  \bibinfo{number}{6} (\bibinfo{date}{jun} \bibinfo{year}{2014}),
  \bibinfo{pages}{577–587}.
\newblock
\showISSN{0362-1340}
\urldef\tempurl%
\url{https://doi.org/10.1145/2666356.2594295}
\showDOI{\tempurl}


\bibitem[Noonan et~al\mbox{.}(2016)]%
        {noonan2016}
\bibfield{author}{\bibinfo{person}{Matt Noonan}, \bibinfo{person}{Alexey
  Loginov}, {and} \bibinfo{person}{David Cok}.}
  \bibinfo{year}{2016}\natexlab{}.
\newblock \showarticletitle{Polymorphic Type Inference for Machine Code}.
\newblock \bibinfo{journal}{\emph{SIGPLAN Not.}} \bibinfo{volume}{51},
  \bibinfo{number}{6} (\bibinfo{date}{jun} \bibinfo{year}{2016}),
  \bibinfo{pages}{27–41}.
\newblock
\showISSN{0362-1340}
\urldef\tempurl%
\url{https://doi.org/10.1145/2980983.2908119}
\showDOI{\tempurl}


\bibitem[Pang et~al\mbox{.}(2021)]%
        {pang2020}
\bibfield{author}{\bibinfo{person}{Chengbin Pang}, \bibinfo{person}{Ruotong
  Yu}, \bibinfo{person}{Yaohui Chen}, \bibinfo{person}{Eric Koskinen},
  \bibinfo{person}{Georgios Portokalidis}, \bibinfo{person}{Bing Mao}, {and}
  \bibinfo{person}{Jun Xu}.} \bibinfo{year}{2021}\natexlab{}.
\newblock \showarticletitle{SoK: All You Ever Wanted to Know About x86/x64
  Binary Disassembly But Were Afraid to Ask}. In \bibinfo{booktitle}{\emph{42nd
  IEEE Symposium on Security and Privacy (SP)}}.
\newblock


\bibitem[Pei et~al\mbox{.}(2021)]%
        {Kexin2021}
\bibfield{author}{\bibinfo{person}{Kexin Pei}, \bibinfo{person}{Jonas Guan},
  \bibinfo{person}{Matthew Broughton}, \bibinfo{person}{Zhongtian Chen},
  \bibinfo{person}{Songchen Yao}, \bibinfo{person}{David Williams-King},
  \bibinfo{person}{Vikas Ummadisetty}, \bibinfo{person}{Junfeng Yang},
  \bibinfo{person}{Baishakhi Ray}, {and} \bibinfo{person}{Suman Jana}.}
  \bibinfo{year}{2021}\natexlab{}.
\newblock \showarticletitle{StateFormer: Fine-Grained Type Recovery from
  Binaries Using Generative State Modeling}. In
  \bibinfo{booktitle}{\emph{Proceedings of the 29th ACM Joint Meeting on
  European Software Engineering Conference and Symposium on the Foundations of
  Software Engineering}} (Athens, Greece) \emph{(\bibinfo{series}{ESEC/FSE
  2021})}. \bibinfo{publisher}{Association for Computing Machinery},
  \bibinfo{address}{New York, NY, USA}, \bibinfo{pages}{690–702}.
\newblock
\showISBNx{9781450385626}
\urldef\tempurl%
\url{https://doi.org/10.1145/3468264.3468607}
\showDOI{\tempurl}


\bibitem[Schuster et~al\mbox{.}(2015)]%
        {schuster2015}
\bibfield{author}{\bibinfo{person}{Felix Schuster}, \bibinfo{person}{Thomas
  Tendyck}, \bibinfo{person}{Christopher Liebchen}, \bibinfo{person}{Lucas
  Davi}, \bibinfo{person}{Ahmad-Reza Sadeghi}, {and} \bibinfo{person}{Thorsten
  Holz}.} \bibinfo{year}{2015}\natexlab{}.
\newblock \showarticletitle{Counterfeit Object-oriented Programming: On the
  Difficulty of Preventing Code Reuse Attacks in C++ Applications}. In
  \bibinfo{booktitle}{\emph{2015 IEEE Symposium on Security and Privacy}}.
  \bibinfo{pages}{745--762}.
\newblock
\urldef\tempurl%
\url{https://doi.org/10.1109/SP.2015.51}
\showDOI{\tempurl}


\bibitem[Shacham(2007)]%
        {rop2007}
\bibfield{author}{\bibinfo{person}{Hovav Shacham}.}
  \bibinfo{year}{2007}\natexlab{}.
\newblock \showarticletitle{The Geometry of Innocent Flesh on the Bone:
  Return-into-Libc without Function Calls (on the X86)}. In
  \bibinfo{booktitle}{\emph{Proceedings of the 14th ACM Conference on Computer
  and Communications Security}} (Alexandria, Virginia, USA)
  \emph{(\bibinfo{series}{CCS '07})}. \bibinfo{publisher}{Association for
  Computing Machinery}, \bibinfo{address}{New York, NY, USA},
  \bibinfo{pages}{552–561}.
\newblock


\bibitem[Snow et~al\mbox{.}(2013)]%
        {Snow2013}
\bibfield{author}{\bibinfo{person}{Kevin~Z. Snow}, \bibinfo{person}{Fabian
  Monrose}, \bibinfo{person}{Lucas Davi}, \bibinfo{person}{Alexandra
  Dmitrienko}, \bibinfo{person}{Christopher Liebchen}, {and}
  \bibinfo{person}{Ahmad-Reza Sadeghi}.} \bibinfo{year}{2013}\natexlab{}.
\newblock \showarticletitle{Just-In-Time Code Reuse: On the Effectiveness of
  Fine-Grained Address Space Layout Randomization}. In
  \bibinfo{booktitle}{\emph{2013 IEEE Symposium on Security and Privacy}}.
  \bibinfo{pages}{574--588}.
\newblock
\urldef\tempurl%
\url{https://doi.org/10.1109/SP.2013.45}
\showDOI{\tempurl}


\bibitem[Team(2015a)]%
        {team2015bctf}
\bibfield{author}{\bibinfo{person}{BlueLotus Team}.}
  \bibinfo{year}{2015}\natexlab{a}.
\newblock \bibinfo{title}{Bctf challenge: bypass vtable read-only checks}.
\newblock
\newblock


\bibitem[Team(2015b)]%
        {rap2015}
\bibfield{author}{\bibinfo{person}{PaX Team}.}
  \bibinfo{year}{2015}\natexlab{b}.
\newblock \showarticletitle{Rap: Rip rop}. In \bibinfo{booktitle}{\emph{Hackers
  2 Hackers Conference (H2HC)}}.
\newblock


\bibitem[Tice et~al\mbox{.}(2014)]%
        {tice2014}
\bibfield{author}{\bibinfo{person}{Caroline Tice}, \bibinfo{person}{Tom
  Roeder}, \bibinfo{person}{Peter Collingbourne}, \bibinfo{person}{Stephen
  Checkoway}, \bibinfo{person}{\'{U}lfar Erlingsson}, \bibinfo{person}{Luis
  Lozano}, {and} \bibinfo{person}{Geoff Pike}.}
  \bibinfo{year}{2014}\natexlab{}.
\newblock \showarticletitle{Enforcing Forward-Edge Control-Flow Integrity in
  GCC \& LLVM}. In \bibinfo{booktitle}{\emph{Proceedings of the 23rd USENIX
  Conference on Security Symposium}} (San Diego, CA)
  \emph{(\bibinfo{series}{SEC'14})}. \bibinfo{publisher}{USENIX Association},
  \bibinfo{address}{USA}, \bibinfo{pages}{941–955}.
\newblock
\showISBNx{9781931971157}


\bibitem[van~der Veen et~al\mbox{.}(2016)]%
        {veen2016}
\bibfield{author}{\bibinfo{person}{Victor van~der Veen}, \bibinfo{person}{Enes
  Göktas}, \bibinfo{person}{Moritz Contag}, \bibinfo{person}{Andre Pawoloski},
  \bibinfo{person}{Xi Chen}, \bibinfo{person}{Sanjay Rawat},
  \bibinfo{person}{Herbert Bos}, \bibinfo{person}{Thorsten Holz},
  \bibinfo{person}{Elias Athanasopoulos}, {and} \bibinfo{person}{Cristiano
  Giuffrida}.} \bibinfo{year}{2016}\natexlab{}.
\newblock \showarticletitle{A Tough Call: Mitigating Advanced Code-Reuse
  Attacks at the Binary Level}. In \bibinfo{booktitle}{\emph{2016 IEEE
  Symposium on Security and Privacy (SP)}}. \bibinfo{pages}{934--953}.
\newblock
\urldef\tempurl%
\url{https://doi.org/10.1109/SP.2016.60}
\showDOI{\tempurl}


\bibitem[Wang et~al\mbox{.}(2015)]%
        {wang2015}
\bibfield{author}{\bibinfo{person}{Minghua Wang}, \bibinfo{person}{Heng Yin},
  \bibinfo{person}{Abhishek~Vasisht Bhaskar}, \bibinfo{person}{Purui Su}, {and}
  \bibinfo{person}{Dengguo Feng}.} \bibinfo{year}{2015}\natexlab{}.
\newblock \showarticletitle{Binary Code Continent: Finer-Grained Control Flow
  Integrity for Stripped Binaries}. In \bibinfo{booktitle}{\emph{Proceedings of
  the 31st Annual Computer Security Applications Conference}} (Los Angeles, CA,
  USA) \emph{(\bibinfo{series}{ACSAC '15})}. \bibinfo{publisher}{Association
  for Computing Machinery}, \bibinfo{address}{New York, NY, USA},
  \bibinfo{pages}{331–340}.
\newblock
\showISBNx{9781450336826}
\urldef\tempurl%
\url{https://doi.org/10.1145/2818000.2818017}
\showDOI{\tempurl}


\bibitem[Xu et~al\mbox{.}(2019)]%
        {confirm2019}
\bibfield{author}{\bibinfo{person}{Xiaoyang Xu}, \bibinfo{person}{Masoud
  Ghaffarinia}, \bibinfo{person}{Wenhao Wang}, \bibinfo{person}{Kevin~W.
  Hamlen}, {and} \bibinfo{person}{Zhiqiang Lin}.}
  \bibinfo{year}{2019}\natexlab{}.
\newblock \showarticletitle{CONFIRM: Evaluating Compatibility and Relevance of
  Control-Flow Integrity Protections for Modern Software}. In
  \bibinfo{booktitle}{\emph{Proceedings of the 28th USENIX Conference on
  Security Symposium}} (Santa Clara, CA, USA)
  \emph{(\bibinfo{series}{SEC'19})}. \bibinfo{publisher}{USENIX Association},
  \bibinfo{address}{USA}, \bibinfo{pages}{1805–1821}.
\newblock
\showISBNx{9781939133069}


\bibitem[Zhang et~al\mbox{.}(2013)]%
        {zhang2013}
\bibfield{author}{\bibinfo{person}{Chao Zhang}, \bibinfo{person}{Tao Wei},
  \bibinfo{person}{Zhaofeng Chen}, \bibinfo{person}{Lei Duan},
  \bibinfo{person}{László Szekeres}, \bibinfo{person}{Stephen McCamant},
  \bibinfo{person}{Dawn Song}, {and} \bibinfo{person}{Wei Zou}.}
  \bibinfo{year}{2013}\natexlab{}.
\newblock \showarticletitle{Practical Control Flow Integrity and Randomization
  for Binary Executables}. In \bibinfo{booktitle}{\emph{2013 IEEE Symposium on
  Security and Privacy}}. \bibinfo{pages}{559--573}.
\newblock
\urldef\tempurl%
\url{https://doi.org/10.1109/SP.2013.44}
\showDOI{\tempurl}


\bibitem[Zhang and Sekar(2013)]%
        {zhangsekar2013}
\bibfield{author}{\bibinfo{person}{Mingwei Zhang} {and} \bibinfo{person}{R.
  Sekar}.} \bibinfo{year}{2013}\natexlab{}.
\newblock \showarticletitle{Control Flow Integrity for COTS Binaries}. In
  \bibinfo{booktitle}{\emph{Proceedings of the 22nd USENIX Conference on
  Security}} (Washington, D.C.) \emph{(\bibinfo{series}{SEC'13})}.
  \bibinfo{publisher}{USENIX Association}, \bibinfo{address}{USA},
  \bibinfo{pages}{337–352}.
\newblock
\showISBNx{9781931971034}


\bibitem[Zhang et~al\mbox{.}(2021)]%
        {Zhang2021}
\bibfield{author}{\bibinfo{person}{Zhuo Zhang}, \bibinfo{person}{Yapeng Ye},
  \bibinfo{person}{Wei You}, \bibinfo{person}{Guanhong Tao},
  \bibinfo{person}{Wen-chuan Lee}, \bibinfo{person}{Yonghwi Kwon},
  \bibinfo{person}{Yousra Aafer}, {and} \bibinfo{person}{Xiangyu Zhang}.}
  \bibinfo{year}{2021}\natexlab{}.
\newblock \showarticletitle{OSPREY: Recovery of Variable and Data Structure via
  Probabilistic Analysis for Stripped Binary}. In
  \bibinfo{booktitle}{\emph{2021 IEEE Symposium on Security and Privacy (SP)}}.
  \bibinfo{pages}{813--832}.
\newblock
\urldef\tempurl%
\url{https://doi.org/10.1109/SP40001.2021.00051}
\showDOI{\tempurl}


\end{thebibliography}

\appendix

\section{Metrics Visualization}

\begin{figure}[h!]
    \centering
    \includegraphics[width=1.0\columnwidth]{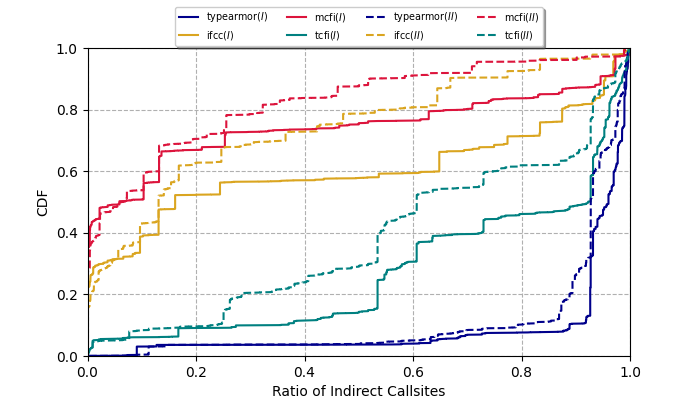}
    \caption{Cumulative Distribution of $RelativeCTR_T$}
    \label{fig:cdft}
\end{figure}

\begin{figure}[h!]
    \centering
    \includegraphics[width=1.0\columnwidth]{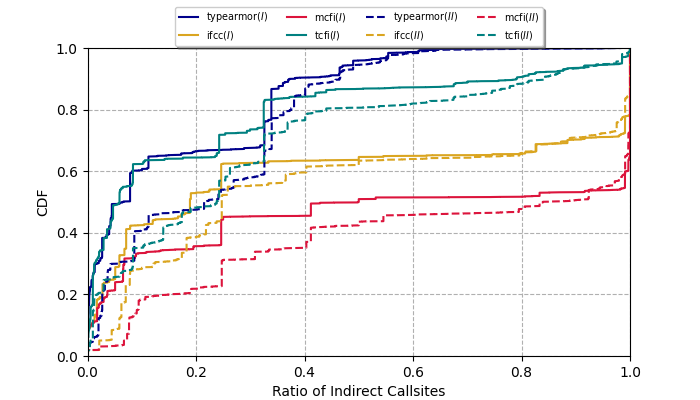}
    \caption{Cumulative Distribution of $RelativeCTR_F$}
    \label{fig:cdff}
\end{figure}

Figures~\ref{fig:cdft} and \ref{fig:cdff} plot the Cumulative Distribution Function (CDF) of the $RelativeCTR$ metrics visualized over all indirect call-sites and all our benchmark programs in both configurations.
The trends in Figure~\ref{fig:cdft} show that the policies on the lower right quadrant of the graph perform better in terms of $RelativeCTR_T$, i.e. correctly identifying reachable target functions (compared to the ground truth). 
TypeArmor, being our most permissive CFI policy, seems to perform better than the other CFI policies we implemented. 

However, Figure~\ref{fig:cdff} displays some contrasting trends.
We now see that TypeArmor doesn't seem to perform better for the same reason, being too permissive. 
After inspecting the two visualizations, we can observe that the $\tau$CFI policy, which uses more relaxed types, shows better promise with moderate levels of $RelativeCTR_T$ and $RelativeCTR_F$ metrics.

\section{Extended Tables}
\label{appendix:extended_results}

Table~\ref{table:spec-opt-stripped:rctr} shows the distribution of the $RelativeCTR$ metrics and Table~\ref{table:spec:ctr} shows the distribution of $CTR$ metrics with Min., Max., Median, 90\textsuperscript{th} percentile, Mean and Standard Deviation values for benchmarks in configurations {\bf I} and {\bf II}.

To visualize the $CTR$ numbers on a more uniform scale, 
Table~\ref{tab:ctf_err} presents the ratio of {\it mean} of the number of reachable call-targets from each call site to the total number of call-targets (functions) in the program.
Thus, the higher ratios for TypeArmor and $\tau$CFI again indicate that they are the more relaxed CFI policies as they permit a greater number of reachable targets, on average, from each call-site.

\begin{table*}[]
\caption{$RelativeCTR_T$ and $RelativeCTR_F$ results for our 4 CFI policies ({\it TypeArmor}, {\it IFCC}, {\it MCFI} and {\it $\tau$cfi}) applied for binaries compiled with symbols (Setting I) and without symbols (Setting II). Mean, Standard Deviation, Minimum, Median, 90\textsuperscript{th} percentile and Maximum aggregate results are displayed for SPEC and Real World benchmarks}
\label{table:spec-opt-stripped:rctr}
\centering
\resizebox{19.5cm}{!}{
\begin{subtable}[]{1.5\textwidth}
\begin{tabular}[]{|l||r|r|r|r|r|r||r|r|r|r|r|r||r|r|r|r|r|r||r|r|r|r|r|r|}
\hline
\multicolumn{1}{|c||}{} & \multicolumn{6}{|c||}{$RelativeCTR_T$ (I)} & \multicolumn{6}{|c||}{$RelativeCTR_F$ (I)} & \multicolumn{6}{|c||}{$RelativeCTR_T$ (II)} & \multicolumn{6}{|c|}{$RelativeCTR_F$ (II)}\\
\hline
 Benchmark           &   Mean &   Std &   Min &   Med &   90thp &   Max &   Mean &   Std &   Min &   Med &   90thp &   Max &   Mean &   Std &   Min &   Med &   90thp &   Max &   Mean &   Std &   Min &   Med &   90thp &   Max \\
\hline
 SPECint             &   0.93 &  0.08 &  0    &  0.93 &    0.99 &     1 &   0.24 &  0.17 &     0 &  0.33 &    0.41 &  0.95  &
 0.92 &  0.10  &  0    &  0.93 &    0.99 &     1 &   0.25 &  0.17 &     0 &  0.33 &    0.46 &  0.95 \\
 SPECfp              &   0.91 &  0.10  &  0.06 &  0.94 &    0.97 &     1 &   0.13 &  0.15 &     0 &  0.07 &    0.22 &  0.77 &
 0.87 &  0.12 &  0.04 &  0.89 &    0.97 &     1 &   0.28 &  0.26 &     0 &  0.18 &    0.64 &  0.86 \\
 nginx      &   0.92 &  0.20  &  0.04 &  0.97 &    0.98 &     1 &   0.03 &  0.10  &     0 &  0.01 &    0.01 &  0.77 &
 0.91 &  0.18 &  0.07 &  0.95 &    0.97 &     1 &   0.19 &  0.12 &     0 &  0.12 &    0.32 &  0.77 \\
 postgresql    &   0.80  &  0.33 &  0.04 &  0.96 &    0.98 &     1 &   0.02 &  0.08 &     0 &  0    &    0    &  0.96 &
 0.75 &  0.30  &  0.06 &  0.87 &    0.91 &     1 &   0.12 &  0.13 &     0 &  0.04 &    0.25 &  0.97 \\
 trafficserver  &   0.93 &  0.08 &  0.11 &  0.93 &    0.98 &     1 &   0.22 &  0.19 &     0 &  0.08 &    0.52 &  0.95  &
 0.93 &  0.08 &  0.11 &  0.93 &    0.98 &     1 &   0.22 &  0.19 &     0 &  0.09 &    0.52 &  0.95 \\
 tor                 &   0.96 &  0.14 &  0.18 &  0.99 &    1    &     1 &   0.12 &  0.18 &     0 &  0.05 &    0.33 &  0.84  &
 0.64 &  0.12 &  0.15 &  0.67 &    0.69 &     1 &   0.29 &  0.17 &     0 &  0.30  &    0.43 &  0.87 \\
 node       &   0.99 &  0.05 &  0.03 &  0.99 &    1    &     1 &   0.05 &  0.10  &     0 &  0.04 &    0.04 &  0.98 &
 0.95 &  0.07 &  0.04 &  0.97 &    1    &     1 &   0.24 &  0.18 &     0 &  0.27 &    0.41 &  0.98 \\
\hline
\end{tabular}
\caption{TypeArmor}\label{table:spec:rtcr:1}
\end{subtable}
}
\\
\resizebox{19cm}{!}{
\begin{subtable}[]{1.5\textwidth}
\begin{tabular}[]{|l||r|r|r|r|r|r||r|r|r|r|r|r||r|r|r|r|r|r||r|r|r|r|r|r|}
\hline
\multicolumn{1}{|c||}{} & \multicolumn{6}{|c||}{$RelativeCTR_T$ (I)} & \multicolumn{6}{|c||}{$RelativeCTR_F$ (I)} & \multicolumn{6}{|c||}{$RelativeCTR_T$ (II)} & \multicolumn{6}{|c|}{$RelativeCTR_F$ (II)}\\
\hline
 Benchmark           &   Mean &   Std &   Min &   Med &   90thp &   Max &   Mean &   Std &   Min &   Med &   90thp &   Max &   Mean &   Std &   Min &   Med &   90thp &   Max &   Mean &   Std &   Min &   Med &   90thp &   Max \\
\hline
 SPECint             &   0.26 &  0.31 &     0 &  0.13 &    0.83 &     1 &   0.45 &  0.38 &     0 &  0.25 &       1 &     1 
 &   0.22 &  0.28 &     0 &  0.13 &    0.83 &  1    &   0.48 &  0.39 &     0 &  0.25 &       1 &     1 \\
 SPECfp              &   0.49 &  0.41 &     0 &  0.61 &    0.91 &     1 &   0.44 &  0.45 &     0 &  0.28 &       1 &     1
 &   0.29 &  0.37 &     0 &  0.12 &    0.93 &  1    &   0.53 &  0.44 &     0 &  0.50  &       1 &     1 \\
 nginx       &   0.68 &  0.45 &     0 &  0.98 &    0.99 &     1 &   0.30  &  0.46 &     0 &  0    &       1 &     1
 &   0.35 &  0.23 &     0 &  0.33 &    0.58 &  0.58 &   0.37 &  0.36 &     0 &  0.10  &       1 &     1 \\
 postgresql     &   0.45 &  0.47 &     0 &  0    &    0.95 &     1 &   0.53 &  0.49 &     0 &  1    &       1 &     1
 &   0.25 &  0.25 &     0 &  0.15 &    0.67 &  1    &   0.52 &  0.39 &     0 &  0.37 &       1 &     1 \\
 trafficserver  &   0.31 &  0.28 &     0 &  0.24 &    0.65 &     1 &   0.39 &  0.42 &     0 &  0.07 &       1 &     1
 &   0.29 &  0.27 &     0 &  0.25 &    0.64 &  1    &   0.40  &  0.42 &     0 &  0.07 &       1 &     1 \\
 tor                 &  0.70  &  0.40  &     0 &  0.91 &    0.96 &     1 &   0.26 &  0.41 &     0 &  0.01 &       1 &     1
 &   0.18 &  0.16 &     0 &  0.15 &    0.37 &  0.59 &   0.51 &  0.34 &     0 &  0.59 &       1 &     1 \\
 node      &   0.74 &  0.34 &     0 &  0.87 &    0.97 &     1 &   0.17 &  0.34 &     0 &  0.03 &       1 &     1
 &   0.31 &  0.31 &     0 &  0.26 &    0.99 &  1    &   0.38 &  0.36 &     0 &  0.22 &       1 &     1 \\
\hline
\end{tabular}
\caption{ifcc}\label{table:spec:rtcr:2}
\end{subtable}
}
\\
\resizebox{19cm}{!}{
\begin{subtable}[]{1.5\textwidth}
\begin{tabular}[]{|l||r|r|r|r|r|r||r|r|r|r|r|r||r|r|r|r|r|r||r|r|r|r|r|r|}
\hline
\multicolumn{1}{|c||}{} & \multicolumn{6}{|c||}{$RelativeCTR_T$ (I)} & \multicolumn{6}{|c||}{$RelativeCTR_F$ (I)} & \multicolumn{6}{|c||}{$RelativeCTR_T$ (II)} & \multicolumn{6}{|c|}{$RelativeCTR_F$ (II)}\\
\hline
 Benchmark           &   Mean &   Std &   Min &   Med &   90thp &   Max &   Mean &   Std &   Min &   Med &   90thp &   Max &   Mean &   Std &   Min &   Med &   90thp &   Max &   Mean &   Std &   Min &   Med &   90thp &   Max \\
\hline
 SPECint             &   0.14 &  0.24 &     0 &  0.06 &    0.52 &     1 &   0.65 &  0.41 &     0 &  0.99 &       1 &     1
 &   0.13 &  0.21 &     0 &  0.06 &    0.52 &  1    &   0.66 &  0.39 &  0    &  0.99 &       1 &     1 \\
 SPECfp              &   0.40  &  0.41 &     0 &  0.33 &    0.94 &     1 &   0.51 &  0.47 &     0 &  0.33 &       1 &     1
 &   0.19 &  0.33 &     0 &  0    &    0.92 &  1    &   0.67 &  0.43 &  0    &  1    &       1 &     1 \\
 nginx      &   0.47 &  0.43 &     0 &  0.66 &    0.99 &     1 &   0.43 &  0.49 &     0 &  0.01 &       1 &     1
 &   0.24 &  0.19 &     0 &  0.26 &    0.61 &  0.61 &   0.72 &  0.34 &  0.15 &  0.96 &       1 &     1 \\
 postgresql    &   0.28 &  0.41 &     0 &  0    &    0.95 &     1 &   0.66 &  0.47 &     0 &  1    &       1 &     1
 &   0.23 &  0.28 &     0 &  0.07 &    0.71 &  1    &   0.76 &  0.35 &  0    &  0.97 &       1 &     1 \\
 trafficserver &   0.10  &  0.14 &     0 &  0    &    0.25 &     1 &   0.51 &  0.40  &     0 &  0.41 &       1 &     1
 &   0.11 &  0.16 &     0 &  0    &    0.26 &  1    &   0.52 &  0.38 &  0    &  0.41 &       1 &     1 \\
 tor                 &   0.49 &  0.41 &     0 &  0.50  &    1    &     1 &   0.32 &  0.45 &     0 &  0    &       1 &     1
 &   0.14 &  0.17 &     0 &  0.08 &    0.40  &  0.59 &   0.76 &  0.34 &  0    &  0.96 &       1 &     1 \\
 node      &   0.64 &  0.37 &     0 &  0.74 &    0.97 &     1 &   0.22 &  0.38 &     0 &  0.05 &       1 &     1
 &   0.28 &  0.32 &     0 &  0.07 &    0.99 &  1    &   0.51 &  0.37 &  0    &  0.40  &       1 &     1 \\
\hline
\end{tabular}
\caption{mcfi}\label{table:spec:rtcr:3}
\end{subtable}
}
\\
\resizebox{19cm}{!}{
\begin{subtable}[]{1.5\textwidth}
\begin{tabular}[]{|l||r|r|r|r|r|r||r|r|r|r|r|r||r|r|r|r|r|r||r|r|r|r|r|r|}
\hline
\multicolumn{1}{|c||}{} & \multicolumn{6}{|c||}{$RelativeCTR_T$ (I)} & \multicolumn{6}{|c||}{$RelativeCTR_F$ (I)} & \multicolumn{6}{|c||}{$RelativeCTR_T$ (II)} & \multicolumn{6}{|c|}{$RelativeCTR_F$ (II)}\\
\hline
 Benchmark           &   Mean &   Std &   Min &   Med &   90thp &   Max &   Mean &   Std &   Min &   Med &   90thp &   Max &   Mean &   Std &   Min &   Med &   90thp &   Max &   Mean &   Std &   Min &   Med &   90thp &   Max \\
\hline
 SPECint             &   0.74 &  0.24 &     0 &  0.80  &    0.93 &  1    &   0.27 &  0.26 &     0 &  0.24 &    0.79 &     1
 &   0.75 &  0.23 &     0 &  0.80  &    0.96 &  1    &   0.27 &  0.26 &     0 &  0.24 &    0.79 &     1 \\
 SPECfp              &   0.89 &  0.16 &     0 &  0.92 &    1    &  1    &   0.16 &  0.29 &     0 &  0.02 &    0.82 &     1
 &   0.71 &  0.23 &     0 &  0.74 &    1    &  1    &   0.28 &  0.32 &     0 &  0.05 &    0.88 &     1 \\
 nginx       &   0.89 &  0.25 &     0 &  0.97 &    0.98 &  1    &   0.03 &  0.16 &     0 &  0    &    0    &     1
 &   0.67 &  0.28 &     0 &  0.49 &    0.97 &  1    &   0.12 &  0.20  &     0 &  0.12 &    0.22 &     1 \\
 postgresql     &   0.74 &  0.32 &     0 &  0.95 &    0.98 &  1    &   0.11 &  0.20  &     0 &  0.05 &    0.42 &     1
 &   0.42 &  0.32 &     0 &  0.26 &    0.92 &  1    &   0.32 &  0.26 &     0 &  0.19 &    0.81 &     1 \\
 trafficserver &  0.48 &  0.24 &     0 &  0.53 &    0.55 &  0.99 &   0.22 &  0.36 &     0 &  0.01 &    0.99 &     1
 &   0.48 &  0.24 &     0 &  0.53 &    0.55 &  0.99 &   0.22 &  0.36 &     0 &  0.01 &    0.99 &     1 \\
 tor          &   0.75 &  0.33 &     0 &  0.96 &    0.99 &  1    &   0.10  &  0.25 &     0 &  0    &    0.42 &     1
 &   0.31 &  0.31 &     0 &  0.21 &    0.69 &  1    &   0.22 &  0.29 &     0 &  0    &    0.67 &     1 \\
 node       &   0.92 &  0.21 &     0 &  0.98 &    1    &  1    &   0.16 &  0.32 &     0 &  0.02 &    0.83 &     1
 &   0.69 &  0.32 &     0 &  0.90  &    1    &  1    &   0.38 &  0.32 &     0 &  0.37 &    0.87 &     1 \\
\hline
\end{tabular}
\caption{tcfi}\label{table:spec:rtcr:4}
\end{subtable}
}
\end{table*}

\begin{table*}[]
\caption{$CTR$ metric results for our 4 CFI policies ({\it TypeArmor}, {\it IFCC}, {\it MCFI} and {\it $\tau$cfi}) applied for binaries compiled with symbols (Setting I) and without symbols (Setting II). Mean, Standard Deviation, Minimum, Median, 90\textsuperscript{th} percentile and Maximum aggregated CTR results results are displayed for SPEC and Real World benchmarks}
\label{table:spec:ctr}
\centering
\resizebox{19cm}{!}{
\begin{subtable}[]{1.4\textwidth}
\begin{tabular}[]{|l||r|r|r|r|r|r||r|r|r|r|r|r||r|r|r|r|r|r|}
\hline
\multicolumn{1}{|c||}{} & \multicolumn{6}{|c||}{Source-CFI} & \multicolumn{6}{|c|}{Binary-CFI (I)} & \multicolumn{6}{|c|}{Binary-CFI (II)}\\
\hline
 Benchmark     &     Mean &      Std &   Min &   Med &   90thp &    Max &     Mean &      Std &   Min &   Med &   90thp &    Max
 & Mean &      Std &   Min &   Med &   90thp &    Max \\
\hline
 SPECint     &  3327.33 &  1917.94 &    15 &  2814 &    6115 &   7880 &  3966.70  &  1843.88 &    15 &  3948 &    6220 &   7880
&  3967.50  &  1848.01 &    19 &  3951 &   6224   &   7880 \\
 SPECfp      &   308.05 &   246.97 &    50 &   274 &     866 &    933 &   307.51 &   225.53 &    23 &   280 &     597 &   1001
&   333.59 &   191.63 &    21 &   296 &    570.2 &   1001 \\
 nginx       &   506.06 &   228.65 &   263 &   411 &     910 &   1209 &   487.85 &   256.97 &    21 &   403 &     892 &   1237
&   570.47 &   278.39 &    48 &   435 &   1009   &   1237 \\
 postgresql    &  6637.11 &  1784.73 &   312 &  6471 &    8046 &  11089 &  5337.80  &  2754.21 &   315 &  5521 &    7816 &  11089
&  5515.54 &  2634.58 &   656 &  5210 &   7775   &  11089 \\
 trafficserver &  3049.97 &   661.98 &   311 &  2877 &    3589 &   6886 &  3882.86 &  1428.00    &   464 &  2893 &    5939 &   6886
&  3905.23 &  1421.62 &   491 &  2921 &   5954   &   6886 \\
 tor           &  2896.82 &  1187.52 &   390 &  3027 &    4436 &   5639 &  3123.42 &  1175.23 &   640 &  3100 &    4455 &   5761
&  2578.18 &   951.86 &   725 &  2948 &   3514   &   5761 \\
 node          & 70251.10  & 26254.10  &  1189 & 63323 &  120968 & 133496 & 73847.80  & 27661.70  &  1429 & 63504 &  121061 & 133496
& 89280.00    & 26653.10  &  2002 & 87139 & 119161   & 133496 \\
\hline
\end{tabular}
\caption{TypeArmor}\label{table:spec:1}
\end{subtable}
}
\resizebox{19cm}{!}{
\begin{subtable}[]{1.45\textwidth}
\begin{tabular}[]{|l||r|r|r|r|r|r||r|r|r|r|r|r||r|r|r|r|r|r|}
\hline
\multicolumn{1}{|c||}{} & \multicolumn{6}{|c||}{Source-CFI} & \multicolumn{6}{|c|}{Binary-CFI (I)} & \multicolumn{6}{|c|}{Binary-CFI (II)}\\
\hline
Benchmark     &     Mean &      Std &   Min &   Med &   90thp &    Max &     Mean &      Std &   Min &   Med &   90thp &    Max
 & Mean &      Std &   Min &   Med &   90thp &    Max \\
\hline
 SPECint       &  1608.22 &  1656.52 &     0 &  1622 &    4015 &  4015 &   606.83 &   966.15 &     0 &   323 &     990 &  3558
&   591.66 &   950.12 &     0 &   334 &     978 &  3558 \\
 SPECfp        &   101.71 &    87.61 &     1 &    95 &     254 &   254 &    78.05 &    84.39 &     0 &    32 &     196 &   282
&    52.72 &    61.76 &     0 &    16 &     152 &   174 \\
 nginx         &   277.97 &   116.97 &     2 &   358 &     358 &   358 &   201.50  &   151.75 &     0 &   181 &     351 &   351
 &   153.71 &    81.38 &     0 &   161 &     229 &   229 \\
 postgresql    &  1825.17 &  1645.56 &     0 &  1123 &    4227 &  4227 &  1233.95 &  1621.13 &     0 &   731 &    4025 &  4025
&  1009.84 &  1049.48 &     0 &   706 &    2959 &  2959 \\
 trafficserver &  1866.46 &   903.06 &     0 &  2432 &    2432 &  2432 &   809.56 &   723.31 &     0 &   626 &    1698 &  1698
&   754.68 &   690.31 &     0 &   640 &    1687 &  1687 \\
 tor           &  923.81 &   649.21 &     1 &   755 &    1814 &  1814 &   730.51 &   616.97 &     0 &   710 &    1654 &  1654
&   365.42 &   295.09 &     0 &   274 &     739 &   739 \\
 node          &  25418.30  & 24289.20  &     0 & 27773 &   59591 & 59591 & 23934.50  & 24164.40  &     0 & 24987 &   58765 & 58765
& 10240.00    & 12128.50  &     0 &  2015 &   28700 & 28700 \\
\hline
\end{tabular}
\caption{ifcc}\label{table:spec:2}
\end{subtable}
}
\resizebox{19cm}{!}{
\begin{subtable}[]{1.45\textwidth}
\begin{tabular}[]{|l||r|r|r|r|r|r||r|r|r|r|r|r||r|r|r|r|r|r|}
\hline
\multicolumn{1}{|c||}{} & \multicolumn{6}{|c||}{Source-CFI} & \multicolumn{6}{|c|}{Binary-CFI (I)} & \multicolumn{6}{|c|}{Binary-CFI (II)}\\
\hline
Benchmark     &      Mean &      Std &   Min &   Med &   90thp &    Max &     Mean &      Std &   Min &   Med &   90thp &    Max
 & Mean &      Std &   Min &   Med &   90thp &    Max \\
\hline
 SPECint       &     1296.44 &  1752.98 &     0 &    81 &    3988 &  3988 &   356.30  &   583.61 &     0 &   323 &     698 &  3029
&  370.67 &   585.27 &     0 &   334 &     699 &  3068 \\
 SPECfp        &     86.55 &    85.79 &     1 &    59 &     250 &   250 &    48.11 &    66.96 &     0 &     5 &     182 &   188
 &   28.00    &    46.96 &     0 &     3 &     145 &   145 \\
 nginx         &   130.45 &   130.80  &     2 &    32 &     319 &   319 &    69.33 &   109.12 &     0 &    21 &     308 &   308
&  142.92 &    91.04 &     0 &   161 &     229 &   229 \\
 postgresql    &   997.45 &  1567.21 &     0 &   130 &    3871 &  3871 &   720.76 &  1233.27 &     0 &    73 &    3662 &  3662
&  932.53 &  1094.99 &     0 &   553 &    2959 &  2959 \\
 trafficserver &   1699.28 &   954.30  &     0 &  2310 &    2310 &  2310 &   229.41 &   318.60  &     0 &    17 &     626 &  1175
&  250.19 &   372.08 &     0 &    17 &     640 &  1464 \\
 tor           &   470.58 &   585.57 &     1 &   202 &    1587 &  1587 &   385.08 &   552.17 &     0 &    50 &    1428 &  1428
&  330.14 &   317.10  &     0 &   274 &     739 &   739 \\
 node          &  17394.30  & 24070.10  &     0 &   771 &   56365 & 56365 & 16165.50  & 23798.90  &     0 &   661 &   55839 & 55839
& 8000.88 & 11631.00    &     0 &   743 &   28700 & 28700 \\
\hline
\end{tabular}
\caption{mcfi}\label{table:spec:3}
\end{subtable}
}
\resizebox{19cm}{!}{
\begin{subtable}[]{1.45\textwidth}
\begin{tabular}[]{|l||r|r|r|r|r|r||r|r|r|r|r|r||r|r|r|r|r|r|}
\hline
\multicolumn{1}{|c||}{} & \multicolumn{6}{|c||}{Source-CFI} & \multicolumn{6}{|c|}{Binary-CFI (I)} & \multicolumn{6}{|c|}{Binary-CFI (II)}\\
\hline
 Benchmark     &   Mean &      Std &   Min &   Med &   90thp &    Max &     Mean &      Std &   Min &   Med &   90thp &    Max
 & Mean &      Std &   Min &   Med &   90thp &    Max \\
\hline
 SPECint       &   2105.19 &  1123.09 &     2 &  2196 &    3301 &  4277 &  2092.90  &  1293.51 &     1 &  2186 &    3866 &  5332
&  2088.97 &  1300.02 &     2 &  2186 &    3869 &   5332 \\
 SPECfp        &   167.72 &   125.53 &     0 &   179 &     332 &   465 &   155.00    &   117.16 &     0 &   163 &     297 &   436
&   120.70  &    84.30  &     0 &   105 &     258 &    486 \\
 nginx         &   366.34 &   291.69 &   148 &   263 &     910 &   983 &   357.17 &   285.79 &   144 &   256 &     891 &   982
&   366.55 &   358.50  &    72 &   320 &    1006 &   1108 \\
 postgresql    &   2415.92 &  1433.94 &    64 &  2484 &    4223 &  7535 &  2372.24 &  1372.63 &    64 &  2508 &    3976 &  7420
&  1585.11 &  2039.42 &    58 &   724 &    4534 &  10095 \\
 trafficserver &   1622.12 &   529.46 &   114 &  1500 &    2286 &  3899 &   990.37 &   977.07 &    39 &   810 &    1043 &  5629
&   991.23 &   978.76 &    39 &   810 &    1044 &   5632 \\
 tor           &   1610.27 &   844.11 &   150 &  1759 &    3027 &  3663 &  1231.44 &   926.35 &   144 &   979 &    3032 &  3692
&   786.90  &  1053.10  &     3 &   363 &    2861 &   3399 \\
 node          &   37759.30  & 21865.70  &   121 & 39706 &   71113 & 71113 & 37679.10  & 22884.40  &   115 & 39974 &   74039 & 74039
& 40666.90  & 42405.30  &     8 & 12830 &  118249 & 118249 \\
\hline
\end{tabular}
\caption{$\tau$cfi}\label{table:spec:4}
\end{subtable}
}
\end{table*}

\begin{table*}[]
    \caption{{\it\{Mean/Total call-targets\}} CTR comparison results of our 4 CFI policies ({\it TypeArmor}, {\it IFCC}, {\it MCFI} and {\it $\tau$cfi})}\label{tab:ctf_err}
    \centering
    \resizebox{1.4\columnwidth}{!}{
    \begin{tabular}{|c|||r|r|r||r|r|r||r|r|r||r|r|r|}
         \hline
         \multicolumn{1}{|c|||}{} & \multicolumn{3}{|c||}{TypeArmor}
         & \multicolumn{3}{|c||}{IFCC} & \multicolumn{3}{|c||}{MCFI}
         & \multicolumn{3}{|c|}{$\tau$CFI}\\
         \hline
        Benchmark & Source & Bin-I & Bin-II & Source & Bin-I & Bin-II & Source & Bin-I & Bin-II & Source & Bin-I & Bin-II\\
        \hline
        SPECint & 0.2134 & 0.2544 & 0.2544 & 0.1031 & 0.0389 & 0.0379 & 0.0831 & 0.0228 & 0.0238 & 0.1350 & 0.1342 & 0.1340 \\
        SPECfp & 0.1316 & 0.1314 & 0.1425 & 0.0434 & 0.0333 & 0.0225 & 0.0370 & 0.0206 & 0.0120 & 0.0716 & 0.0662 & 0.0516 \\
        nginx & 0.2490 & 0.2486 & 0.2697 & 0.0822 & 0.0631 & 0.0426 & 0.0700 & 0.0389 & 0.0226 & 0.1356 & 0.1253 & 0.0976 \\
        postgresql & 0.0278 & 0.0277 & 0.0301 & 0.0092 & 0.0070 & 0.0048 & 0.0078 & 0.0043 & 0.0025 & 0.0151 & 0.0140 & 0.0109 \\
        trafficserver & 0.0447 & 0.0447 & 0.0484 & 0.0148 & 0.0113 & 0.0077 & 0.0126 & 0.0070 & 0.0041 & 0.0244 & 0.0225 & 0.0175 \\
        tor & 0.0535 & 0.0534 & 0.0579 & 0.0177 & 0.0135 & 0.0092 & 0.0150 & 0.0084 & 0.0049 & 0.0291 & 0.0269 & 0.0210 \\
        node & 0.0023 & 0.0023 & 0.0025 & 0.0008 & 0.0006 & 0.0004 & 0.0006 & 0.0004 & 0.0002 & 0.0013 & 0.0012 & 0.0009 \\
        \hline
    \end{tabular}
    }
\end{table*}

\begin{table*}[]
    \caption{Call-sites with zero reachable call-targets}\label{tab:zero_targets}
    \centering
    \resizebox{\columnwidth}{!}{
    \begin{tabular}{|c|||r|r|r||r|r|r||r|r|r|}
         \hline
         \multicolumn{1}{|c|||}{}
         & \multicolumn{3}{|c||}{IFCC} & \multicolumn{3}{|c||}{MCFI}
         & \multicolumn{3}{|c|}{$\tau$CFI}\\
         \hline
        Benchmark & Source & Bin-I & Bin-II & Source & Bin-I & Bin-II & Source & Bin-I & Bin-II\\
        \hline
        SPECint & 4 & 692 & 971 & 62 & 4233 & 3320 & 0 & 0 & 0 \\
        SPECfp & 0 & 152 & 145 & 0 & 401 & 499 & 31 & 31 & 31 \\
        nginx & 0 & 9 & 9 & 0 & 23 & 20 & 0 & 0 & 0 \\
        postgresql & 16 & 452 & 256 & 208 & 1249 & 518 & 0 & 0 & 0 \\
        trafficserver & 1 & 114 & 123 & 2 & 1123 & 927 & 0 & 0 & 0 \\
        tor & 0 & 13 & 13 & 0 & 18 & 35 & 0 & 0 & 0 \\
        node & 3 & 226 & 308 & 17 & 559 & 736 & 0 & 0 & 0 \\
        \hline
    \end{tabular}
    }
\end{table*}

\section{Call-sites with zero call-targets}
\label{appendix:zero_calltargets}
We currently implement a strict and pedantic version of each CFI policy (as done in LLVM-CFI). A strict implementation of CFI policies occasionally results in the ``odd'' situation where a few call-sites have zero legal targets.  Table~\ref{tab:zero_targets} shows the number of call-sites with zero legal targets in the CFI policies we implement and for all three of our benchmark configurations. In practice, we expect that most policies will have a fail-safe to handle such cases. We plan to extend our policies with additional heuristics to handle such cases in the future.

\end{document}